\documentclass[usenatbib,referee,twocolumn]{mn2e}
\usepackage{graphicx}
\usepackage{amssymb}
\usepackage{times}

\def\fdisc{f_{\rm disc}}
\def\thetabar{\theta_{\rm bar}}
\def\tauogle{\tau_{\rm OGLE}}
\def\tauoglelos{\tau_{\rm OGLE\ LOS}}
\def\taumacho{\tau_{\rm MACHO}}
\def\taumacholos{\tau_{\rm MACHO\ LOS}}
\def\taumoa{\tau_{\rm MOA}}
\def\taumoalos{\tau_{\rm MOA\ LOS}}
\def\taueros{\tau_{\rm EROS}}
\def\tauflat{\tau_{\rm flat}}
\def\Iogle{I_{\rm OGLE}}
\def\Reros{R_{\rm EROS}}

\def\B-Reros{(B - R)_{\rm EROS}}
\def\Rclump{R_{\rm EROS,\ clump}}

\def\Iclump{I_{\rm 0,\ clump}}

\newcommand\apj{{ApJ}}     
\newcommand\aap{{A\&A}}    
\newcommand\mnras{{MNRAS}} 

\title[]{Microlensing optical depth as a function of source apparent magnitude}
\author[Wood]{Alexander Wood
\thanks{{Email: A.A.Wood@postgrad.manchester.ac.uk}} \\
Jodrell Bank Observatory, University of Manchester, Macclesfield, Cheshire SK11 9DL
}

\date{Accepted ........ Received ........; in original form ........}
\pubyear{2007}

\begin{document}
\maketitle

\begin{abstract}

Measurements of the microlensing optical depth $\tau$ towards the Galactic bulge 
appear to depend on the method used to obtain them. Those values based on the lensing 
of red clump giants (RCGs) appear to be significantly lower than those based on the 
lensing of all stars along the line of sight. This discrepancy is still not understood. 
Through Monte Carlo simulations, it is found that the discrepancy cannot be explained 
by a dependance on the flux limits of the two methods. The optical depth is expected 
to be generally constant as a function of source apparent magnitude for $I_0 \gtrsim 
13.0$, except in the range $13.5 \lesssim I_0 \lesssim 15.5$. Here many RCGs are 
detected, causing a significant oscillation in $\tau$. The amplitude of this 
oscillation is a function of the inclination angle of the Galactic bar, $\thetabar$, 
which may thus be constrained. A further constraint comes from a similar dependance 
of $\tau$ with $\thetabar$: combining the predicted trends with the measured values 
provides $1 \sigma$ upper limits, which exclude the large bar angles recently reported 
by the GLIMPSE and EROS surveys. The latest survey data from EROS-2 appear to show the 
predicted $\tau$ oscillation, though currently at a low significance. However, a 
further sign comes from EROS-2 event counts, which show a clear skew towards fainter 
magnitudes.
\end{abstract}


\section{Introduction}
\label{sec:intro}

Thousands of microlensing events in our Galaxy have so far been discovered by OGLE 
(e.g. \citealt{Woz01}), MACHO (e.g. \citealt{Tho05}), MOA (e.g. \citealt{Sum03}) and 
EROS (e.g. \citealt{Afo03}), with many more detected every year\footnote{Real-time 
alerts are available online, e.g. the OGLE Early Warning System: 
www.astrouw.edu.pl/$\sim$ogle/ogle3/ews/ews.html}. One of the most important 
measurements that can be made from these observations is of the optical depth, $\tau$ 
-- the probability of seeing a microlensing event. However, the measured value appears 
to depend strongly on the method used to obtain it.

\citet{Pop05} reported $\taumacho = 2.17^{+0.47}_{-0.38} \times 10^{-6}$ at $(l, b) = 
(1.50^\circ, -2.68^\circ)$, and more recently from the OGLE-II survey, \citet{Sum06} 
found $\tauogle = 2.55^{+0.57}_{-0.46} \times 10^{-6}$ at $(l, b) = (1.16^\circ, 
-2.75^\circ)$. The MACHO value was based on the lensing of 42 red clump giants (RCGs), 
and used standard photometric fitting. The OGLE analysis instead used difference image 
analysis (DIA), but was similarly based on the lensing of 32 red giants, red super 
giants and RCGs, and obtained an optical depth consistent with the previous MACHO 
result. However, both these values are significantly lower than two other recent 
measurements, which were based on the lensing of all stars. Using 28 DIA events, 
\citet{Sum03} found $\taumoa = 3.36^{+1.11}_{-0.81} \times 10^{-6}\, 
[0.77 / (1 - \fdisc)]$, where $\fdisc$ is the contribution from disc sources -- the 
coordinates of this value are given in \citet{Sum06}: $(l, b) = (3.0^\circ, 
-3.8^\circ)$. \citet{Alc00} had previously found $\tau = 3.23^{+0.52}_{-0.50} \times 
10^{-6}\, [0.75 / (1 - \fdisc)]$ at $(l, b) = (2.68^\circ, -3.35^\circ)$, from 99 MACHO 
DIA events. The latest measurement comes from the EROS-2 survey of bulge RCGs, which 
yielded 120 events: \citet{Ham06} give the trend $\taueros = (1.62 \pm 0.23)\, 
{\rm exp}[-a(|b| - 3\ {\rm deg})] \times 10^{-6}$, where $a = (0.43 \pm 0.16\ 
{\rm deg}^{-1})$, in the latitude range $1.4^\circ < |b| < 7.0^\circ$. This agrees well 
with previous EROS values, and with the recent MACHO and OGLE-II measurements.

The question naturally arises as to why the RCG-based optical depths appear to be lower 
than those from all stars. One possibility is a dependence on the flux limits of the 
two methods. RCGs are bright; the latter method will include much fainter stars, and so 
probe sources at greater distances, which will have a higher optical depth 
(\citealt{Sta95}).

This potential explanation of the discrepancy is investigated using Monte Carlo 
simulations of Galactic microlensing, and the optical depth as a function of source 
apparent magnitude is then predicted. The model Galaxy is barred, and in light of 
observations by the \emph{Spitzer Space Telescope} (\emph{SST}), that support a bar 
inclination angle much larger than suggested by previous studies (see 
\S\ref{sec:models}), the effect on the expected $\tau$ of changing the bar angle is 
determined. Combining these results with the observed optical depths, upper limits are 
placed on the bar angle. \S\ref{sec:model} describes the model. The results and 
discussion are presented in \S\ref{sec:results}: the model results are given in 
\S\ref{sec:results_model}, and comparisons are made with the latest EROS-2 data in 
\S\ref{sec:results_eros}. A summary and conclusions follow in \S\ref{sec:conclusions}.

\section{The Model}
\label{sec:model}

\subsection{Bulge and disc mass models}
\label{sec:models}

The mass models and parameters of the Galactic bulge (bar) and disc are as described 
in \citet{WM05}. They are based on those of \citet{HG03}, who empirically normalised 
the G2 bulge model of \citet*[table 1]{Dwe95} with \emph{Hubble Space Telescope} star 
counts, and extended the local disc model of \citet{Zhe01} to the whole disc. 
\citeauthor{Dwe95} tested a series of models against images of the Galactic bulge from 
the \emph{Cosmic Background Explorer} (\emph{COBE}) satellite, and found their G2 model 
to provide one of the best fits. This model bar extends from 3--13 kpc and is inclined 
to the Galactic centre line of sight (LOS) at an angle of $\thetabar = 13.4^\circ$. 
\citet{Ger02} states that physical models can be found for the \emph{COBE} bar with 
angles in the range $15^\circ \lesssim \thetabar \lesssim 35^\circ$, and many studies 
assume $\thetabar \approx 20^\circ$. However, more recent data from GLIMPSE (Galactic 
Legacy Infrared Mid-Plane Survey Extraordinaire), using the \emph{SST}, support a much 
larger value of $(44 \pm 10)^\circ$ (\citealt{Ben05}), while from EROS-2, \citet{Ham06} 
report $\thetabar = (49 \pm 8)^\circ$, which is consistent with original OGLE-I results 
(\citealt{Sta94}). Hence, predictions are also made here for \citeauthor{Dwe95}'s E2 
model, which has the largest bar angle of their models: $\thetabar = 41.3^\circ$.

\subsection{Source population}
\label{sec:sources}

The expected $\tauogle$, $\taumacho$ and $\taumoa$ are to be calculated. Therefore for 
each LOS, the apparent magnitude distribution of the model sources must match the 
observed distribution. \citet{Sum04} fitted the $I$-band stellar distributions in 48 
OGLE-II Galactic bulge fields with the power-law plus Gaussian luminosity function
\begin{equation}
\phi_I(I) = p_0 10^{p_1 I} + p_2 \exp \left[ -\frac{(I - \langle I \rangle_{\rm RC})^2}{2 \sigma_{I, \rm RC}^2} \right],
\label{eq:powgauss}
\end{equation}
where $p_0$, $p_1$, $p_2$ and $\sigma_{I, \rm RC}$ are free parameters, and 
$\langle I \rangle_{\rm RC}$ is measured as described in his paper. The power-law part 
contains red giants and bright main-sequence stars, which lie throughout the bar. The 
Gaussian component consists of RCGs, which in the model here are more concentrated in 
the central part of the bulge, occupying the region 6--10 kpc. (This concentration is 
found to improve the match to the observed magnitude distributions, and is not 
unreasonable, as RCGs are older, evolved stars, and hence more likely to exist only in 
more densely populated regions).

\citet{Sum04} thus provides, for each of these fields, an observed distribution of 
apparent magnitude. The positions of these fields are listed in table 1 of 
\citet{Uda02}. 

The fields closest to the OGLE, MACHO (RCG) and MOA lines of sight are selected, as 
shown in Table \ref{tab:coords}. The MACHO (DIA) LOS is not considered, as explained 
below.

\begin{table}
\centering
\begin{tabular}{lccc}\hline

  & $l$ ($^\circ$) & $b$ ($^\circ$) & Angular separation ($^\circ$) \\ \hline

  OGLE     & 1.16 & $-2.75$ &      \\
  Field 34 & 1.35 & $-2.40$ & 0.40 \\ \\
  MACHO    & 1.50 & $-2.68$ &      \\
  Field 20 & 1.68 & $-2.47$ & 0.28 \\ \\
  MOA      & 3.0  & $-3.8$  &      \\
  Field 36 & 3.16 & $-3.20$ & 0.6  \\ \hline

\end{tabular}
\caption{Selection of the OGLE-II Galactic bulge fields that are closest to the lines 
  of sight of the OGLE, MACHO (RCG) and MOA optical depth measurements.}
\label{tab:coords}
\end{table}

As described in \S\ref{sec:opdepth}, the apparent magnitude of each source is 
calculated by first assigning it an absolute magnitude, and then correcting for its 
distance. Hence for each LOS a separate model distribution of absolute magnitude is 
required that will, with distance corrections, reproduce the observed distribution of 
apparent magnitude. Of course in reality the absolute magnitude distribution should be 
virtually the same for each direction in Table \ref{tab:coords}, since over these small 
angular separations the mass function is expected to vary little. Here, the artificial 
absolute magnitude distributions are only used as a means to ensure the model 
distributions of apparent magnitude match those observed for each direction. It is 
assumed that the forms of the two distributions are the same, i.e. a power-law plus 
Gaussian. For each of the three OGLE-II fields listed in Table \ref{tab:coords}, an 
appropriate absolute magnitude distribution can easily be generated, by suitably 
adjusting the (extinction-corrected) fitted parameters of the apparent magnitude 
distribution found by \citet{Sum04}. (These parameters are not given in \citet{Sum04}, 
and are provided by T. Sumi, private communication).

\citet*[table 5]{Sum03} and \citet*[table 4]{Sum06} list the extinction-corrected 
$I$-band apparent magnitudes of all the MOA and OGLE sources used in their respective 
measurements of $\tau$. The minimum and maximum magnitudes given in each case are taken 
to define ranges of detectable apparent magnitudes. \citet{Pop05}, in their table 2, 
provide uncorrected $V$-band apparent magnitudes. Since these MACHO sources are all 
RCGs, their apparent magnitudes are converted to $I$-band using the following relation 
for RCGs from \citet{Sum03}:
\begin{equation}
I = (1.45 \pm 0.12)\, (V - I) + 12.7.
\label{eq:rcgs}
\end{equation}
MACHO source extinction is then accounted for by simply shifting the minimum and 
maximum MACHO magnitudes by the mean $A_I$ for the corresponding OGLE-II field, as 
given in table 3 of \citet{Sum04}: for field 20, $A_I$ = 0.951. (Strictly speaking the 
extinction should instead be calculated for each source individually, but this is 
neglected as in the region occupied by the model RCGs, $A_I$ varies from the mean value 
by $\lesssim$ 0.05 mag -- a negligible amount). Although \citet{Alc00} also list the 
apparent magnitudes of the MACHO sources used in their DIA measurement, these are in 
$V$-band and do not consist of only RCGs. Therefore an $I$-band magnitude range cannot 
be reliably defined for the model. The defined ranges of detectable, 
extinction-corrected apparent magnitudes $I_0$ for OGLE, MACHO and MOA are given in 
Table \ref{tab:mag_ranges}.

\begin{table}
\centering
\begin{tabular}{lcc}\hline

        & $I_{\rm 0, min}$ & $I_{\rm 0, max}$ \\ \hline
  OGLE  & 12.1             & 15.3             \\
  MACHO & 13.9             & 16.2             \\
        & ($V$ = 16.37)    & ($V$ = 20.19)    \\
  MOA   & 13.6             & 20.8             \\ \hline

\end{tabular}
\caption{Defined ranges of detectable, extinction-corrected apparent magnitudes for 
  OGLE, MACHO and MOA.} 
\label{tab:mag_ranges}
\end{table}

\begin{table*}
\centering
\newcommand\T{\rule{0pt}{2.6ex}}
\begin{tabular}{lcccc}\hline

  & ($l, b$) ($^\circ$) & $\tau_{\rm obs}$ & $\tau_{\rm model, G2}$ & $\tau_{\rm model, E2}$ \\
  & & $(\times 10^{-6})$ & $(\times 10^{-6})$ & $(\times 10^{-6})$ \\ \hline

  OGLE  \T & (1.16, $-2.75$) & $2.55^{+0.57}_{-0.46}$                          & 2.14 & 1.57 \\
  MACHO \T & (1.50, $-2.68$) & $2.17^{+0.47}_{-0.38}$                          & 2.19 & 1.61 \\
  MOA   \T & (3.0, $-3.8$)   & $3.36^{+1.11}_{-0.81}$ ($2.59^{+0.84}_{-0.64}$) & 1.38 & 1.01 \\ \hline

\end{tabular}
\caption{The expected G2 optical depths agree well with those reported by OGLE and 
  MACHO, but not with MOA's values. (The numbers in parentheses are without the DIA 
  correction for disc sources -- see text).}
\label{tab:opdepths}
\end{table*}

\subsection{Optical depth}
\label{sec:opdepth}

The expected $\tauogle$, for example, can now be calculated as follows. First a 
distance is chosen for a given source along the OGLE LOS. It is assigned an absolute 
magnitude using the artificial distribution constructed for the nearest OGLE-II field, 
\#34. The source's apparent magnitude is then calculated by accounting for its distance.

If this apparent magnitude falls within the defined range of magnitudes detectable by 
OGLE, the source is included in the calculation of $\tau$ (using equation (5) of 
\citealt{WM05}). This process is then repeated for many sources. The expected 
$\taumacho$ and $\taumoa$ are similarly calculated.

\section{Results and discussion}
\label{sec:results}

\subsection{Model results}
\label{sec:results_model}

\begin{figure}
\centering
\includegraphics[width = 7.5cm]{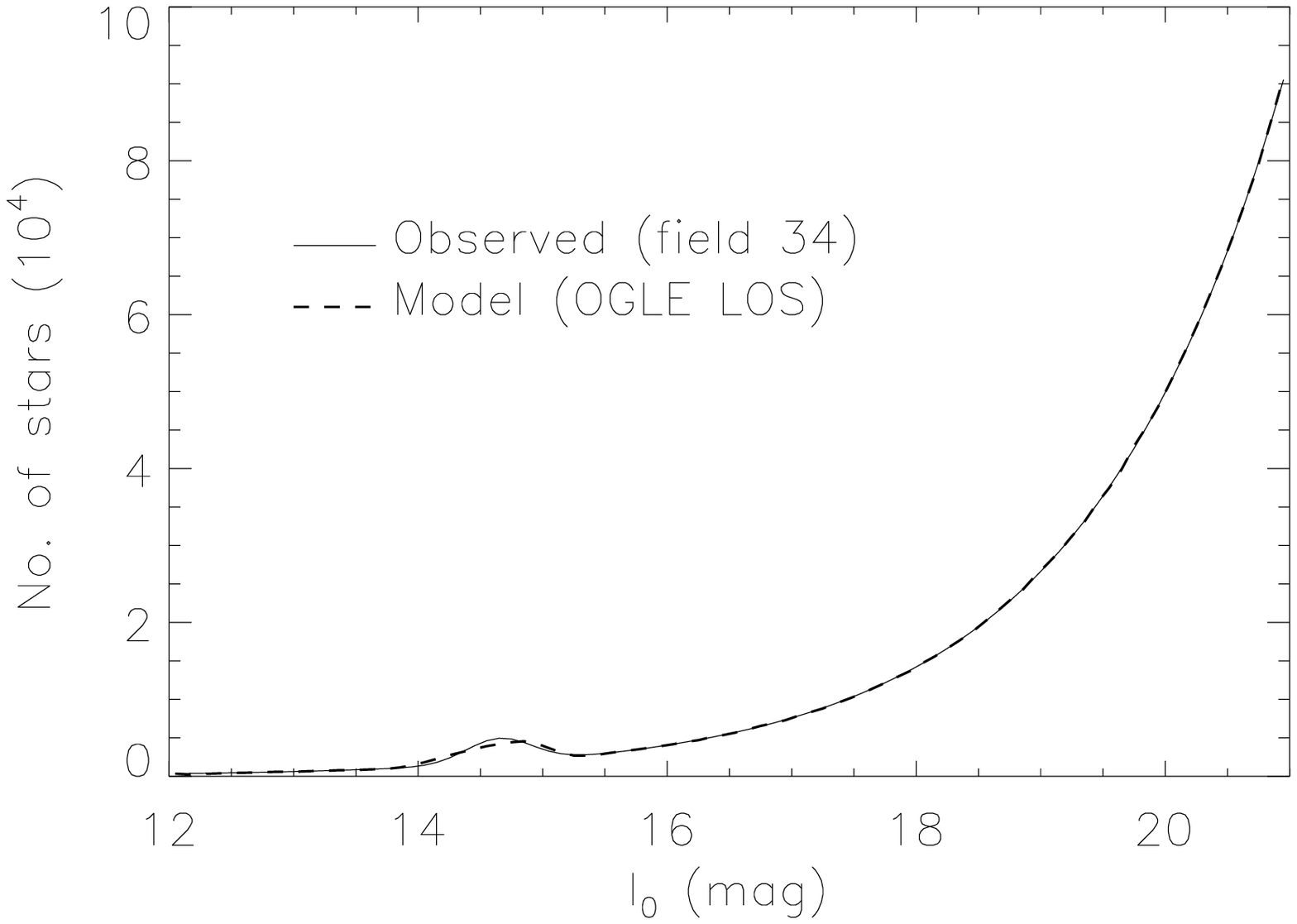}
\includegraphics[width = 7.5cm]{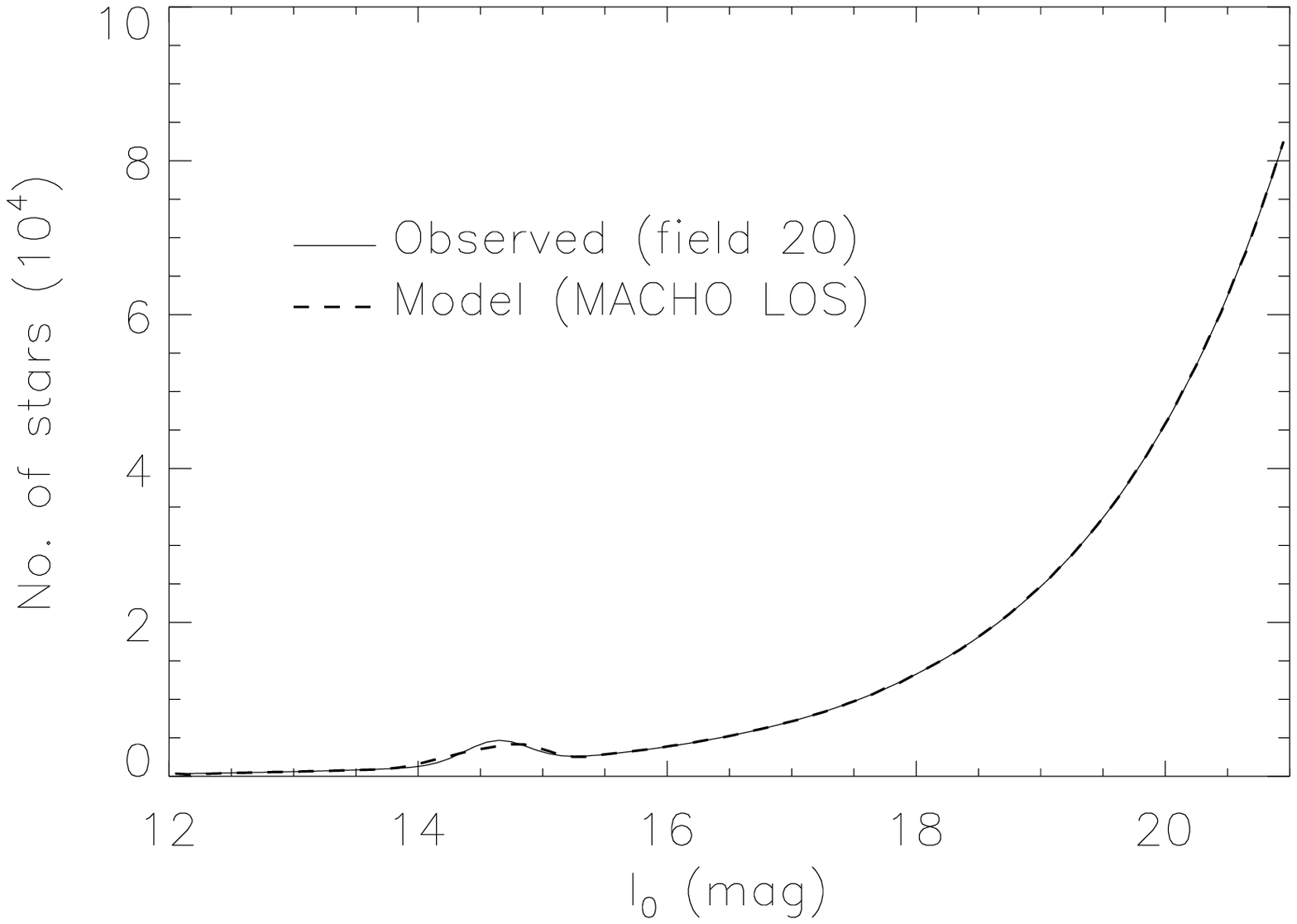}
\includegraphics[width = 7.5cm]{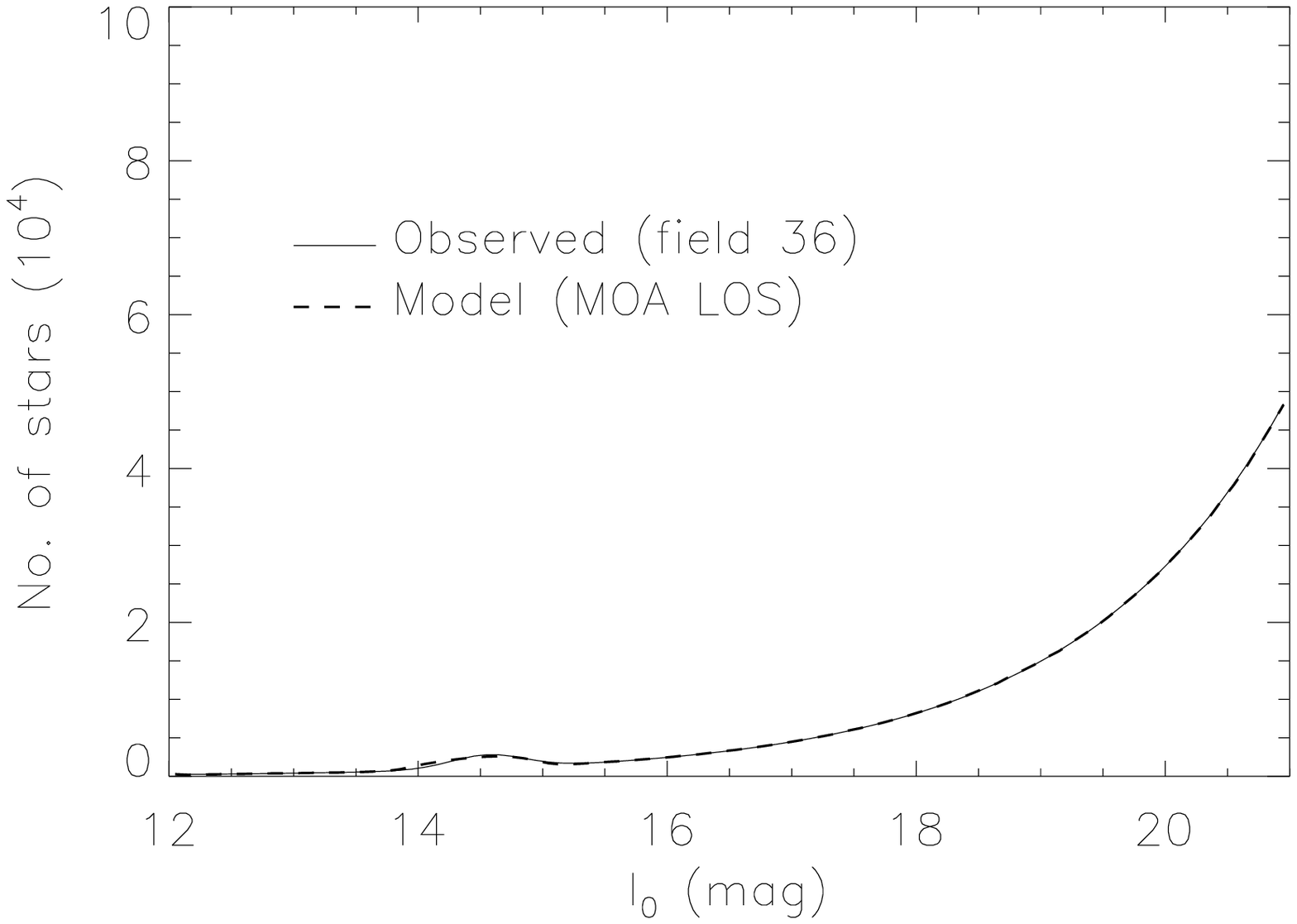}
\caption{(Top, middle, bottom) panel: apparent magnitude distributions, for stars 
  observed in OGLE-II field (34, 20, 36) by \citet{Sum04}, and model stars along the 
  (OGLE, MACHO, MOA) LOS. These plots are for the G2 model -- there is negligible 
  difference with E2. The model curves are normalised to the same area as the 
  corresponding observed curves.}
\label{fig:freq_sources}
\end{figure}

Fig. \ref{fig:freq_sources} shows that the model well reproduces the observed 
distributions of apparent magnitude for fields 20, 34 and 36. Table \ref{tab:opdepths} 
shows that the expected values of $\tauogle$ and $\taumacho$ also agree well with the 
observed values, for the G2 model. However, the expected $\taumoa$ lies $\sim$$2.4 
\sigma$ below the reported value. As the MOA measurement is sensitive to \emph{all} 
sources along the LOS, a correction was applied to account for disc sources. This is 
expressed by the $\fdisc$ term in the $\tau$ measurements quoted in \S\ref{sec:intro}. 
Such adjustments typically raise $\tau$ by $\sim$25 per cent. Note that the model 
underpredicts $\taumoa$ by a much greater margin, hence the disagreement cannot be 
attributed to the correction applied by MOA. It therefore appears that the discrepancy 
in the survey measurements cannot be simply explained by a dependance on their 
different flux limits.

\begin{figure}
\centering
\includegraphics[width = 7.5cm]{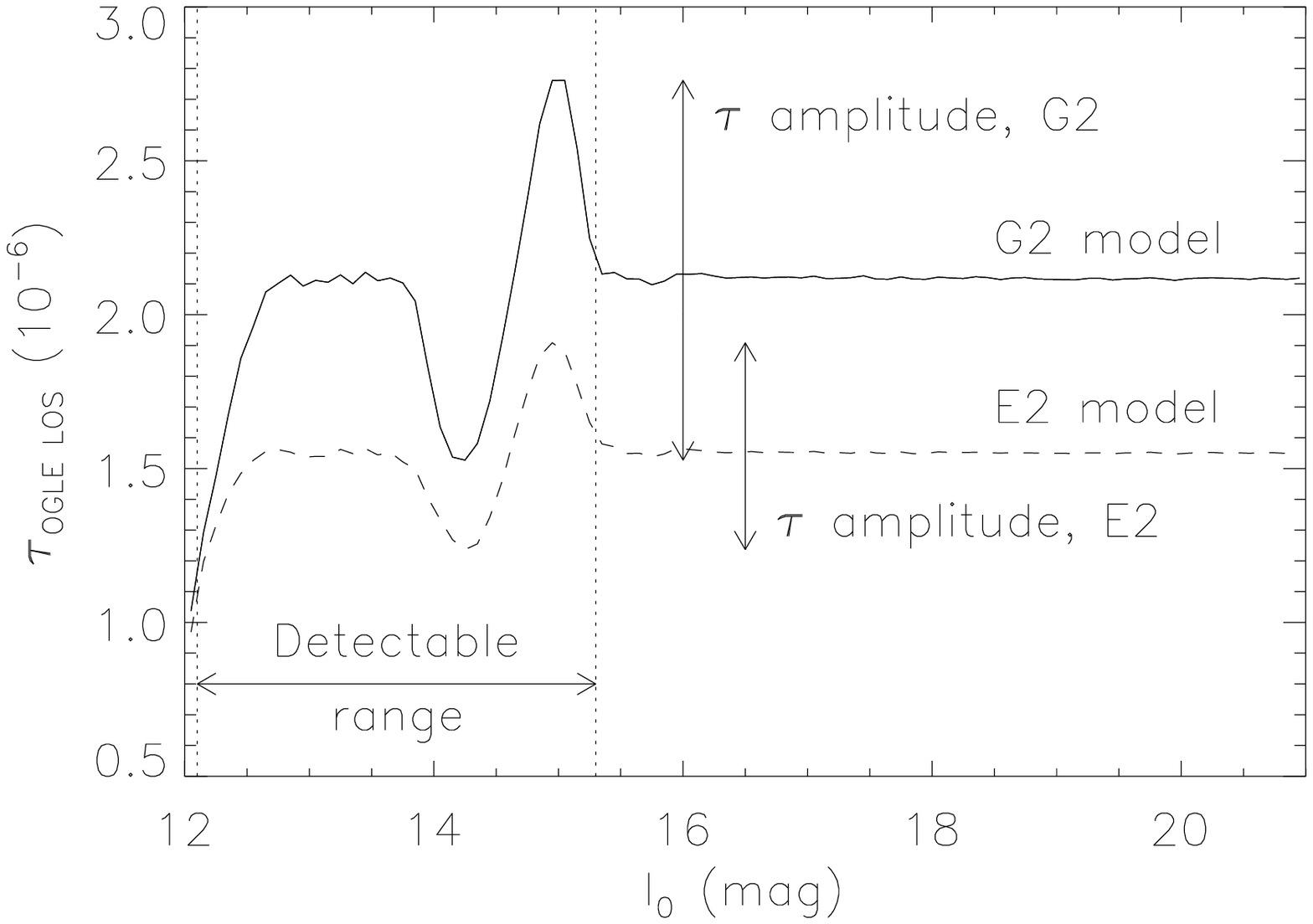}
\includegraphics[width = 7.5cm]{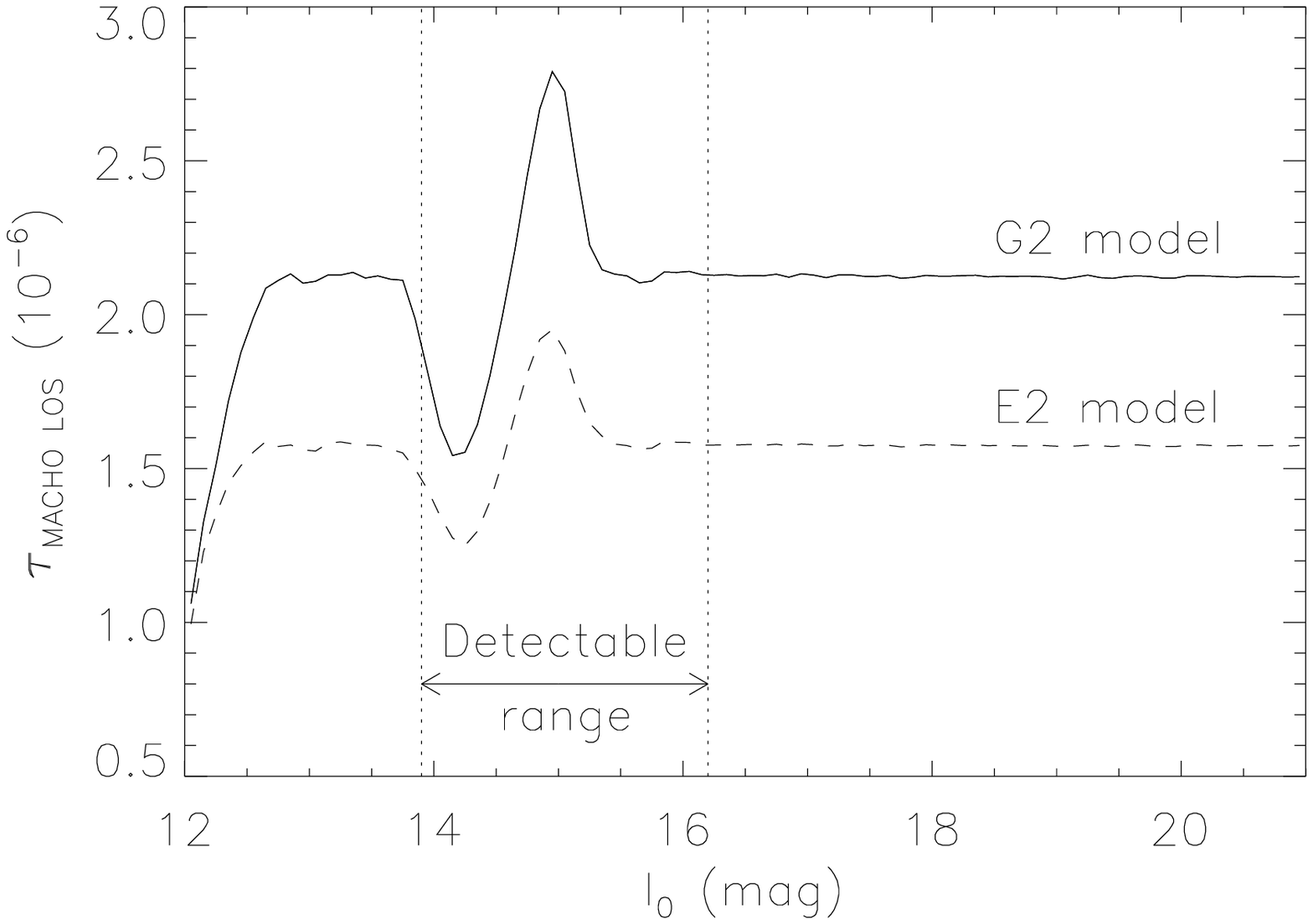}
\includegraphics[width = 7.5cm]{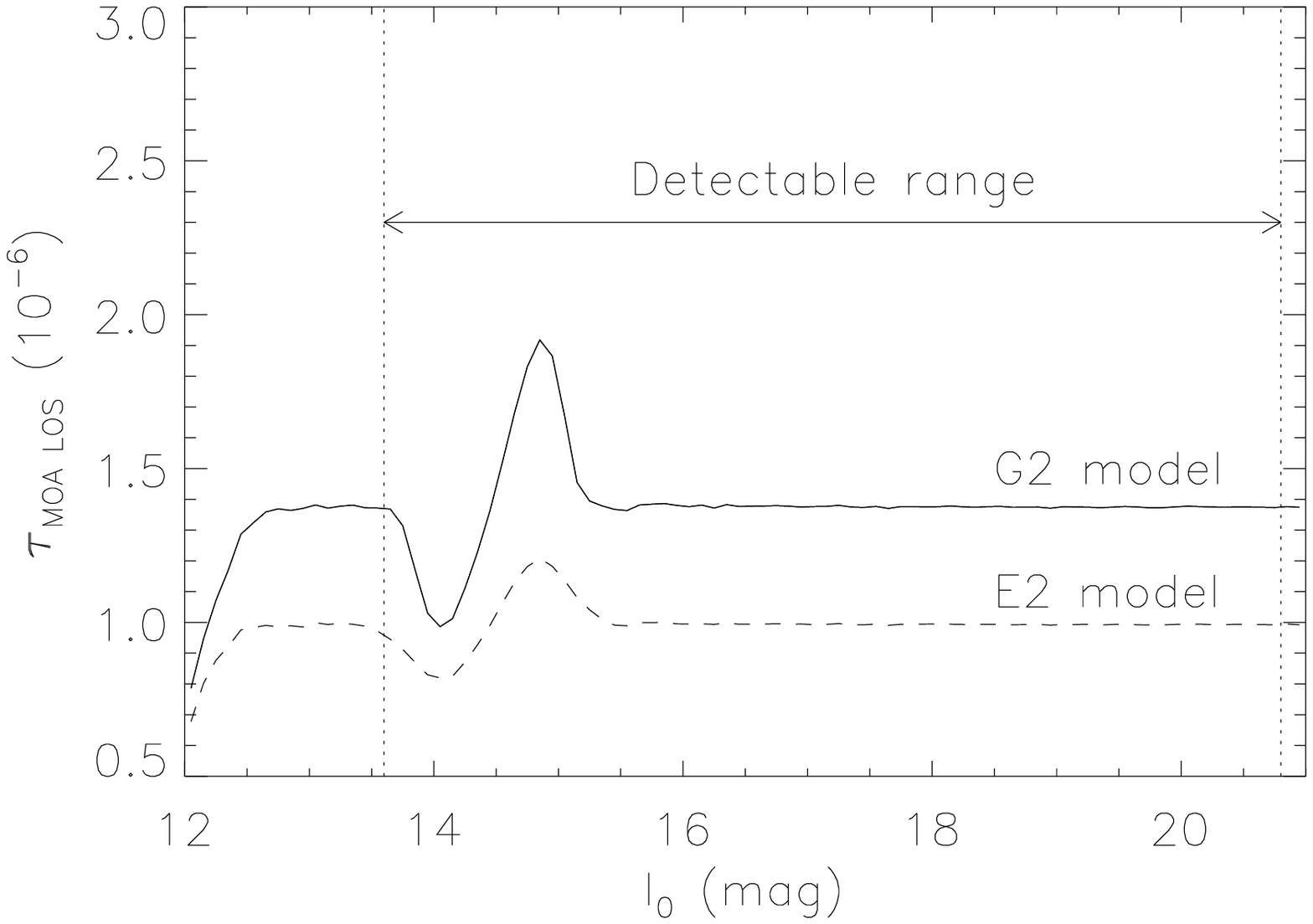}
\caption{(Top, middle, bottom) panel: expected ($\tauoglelos$, $\taumacholos$, 
  $\taumoalos$) as a function of source apparent magnitude, for the G2 and E2 models. 
  The detectable magnitude ranges given in Table \ref{tab:mag_ranges} are shown. In the 
  top panel, the amplitude of the $\tauoglelos$ oscillation (see text) is indicated for 
  both the G2 and E2 models.}
\label{fig:magtau}
\end{figure}

However, there may be other ways in which $\tau$ depends on the source flux. So far 
the predicted optical depths have been calculated by summing over all the source stars 
whose apparent magnitudes fall within specified ranges. By repeating this process for 
many small bins of $I_0$, $\tau$ can be predicted as a function of $I_0$. This is 
plotted in Fig. \ref{fig:magtau}, for the OGLE, MACHO and MOA coordinates 
($\tauoglelos$, $\taumacholos$ and $\taumoalos$, respectively). The detectable 
magnitude ranges given in Table \ref{tab:mag_ranges} are also shown. For each LOS the 
absolute expected value of $\tau$ is higher for the G2 bar than the E2, but its trend 
with magnitude is similar. These trends are explained as follows.

$\tau$ increases rapidly over the range $12 \lesssim I_0 \lesssim 13$. Almost all 
sources of magnitude $\sim$12 will be on the near side on the bulge, so as $I_0$ 
increases fainter and more distant stars, with higher optical depths, come into view. 
For $I_0 \gtrsim 15.5$, $\tau$ is approximately constant. This is because the power-law 
part of the source magnitude distribution spans a wide range of $I_0$. Hence these 
stars can be either bright or faint whether they are near or far, and thus will show 
little or no correlation between apparent magnitude and distance. So, when calculating 
the average $\tau$ for a given apparent magnitude, the lower optical depth of the 
closer stars is balanced by the higher $\tau$ of those more distant.

In comparison, the Gaussian (RCG) part of the source distribution covers only a very 
narrow range of absolute magnitudes. The RCGs' distribution in apparent magnitude will 
be broader, due to variations in their distance, but as they are more concentrated in 
the centre of the bulge, this broadening is not great. Therefore the vast majority of 
RCGs will lie within a small range of apparent magnitude, and hence show a strong 
correlation between apparent magnitude and distance. At $I_0 \approx 14$ we see many 
RCGs, and they greatly outnumber the other sources. Most of the RCGs at this magnitude 
lie on the near side of the bulge, and $\tau$ is lower. As $I_0$ increases, the average 
distance of the RCGs (and so of all sources) being observed shifts towards the far side 
of the bulge, and $\tau$ increases. As $I_0$ becomes fainter still, $\gtrsim 15$, we 
see fewer and fewer RCGs, and the average distance of all the observed sources moves 
back towards the centre of the bulge, where it then remains, and $\tau$ becomes 
approximately constant. The amplitude of this oscillation in $\tau$ (hereafter the 
\emph{$\tau$ amplitude}) caused by the RCGs along the OGLE LOS is indicated in the top 
panel of Fig. \ref{fig:magtau}.

\begin{figure}
\centering
\includegraphics[width = 7.5cm]{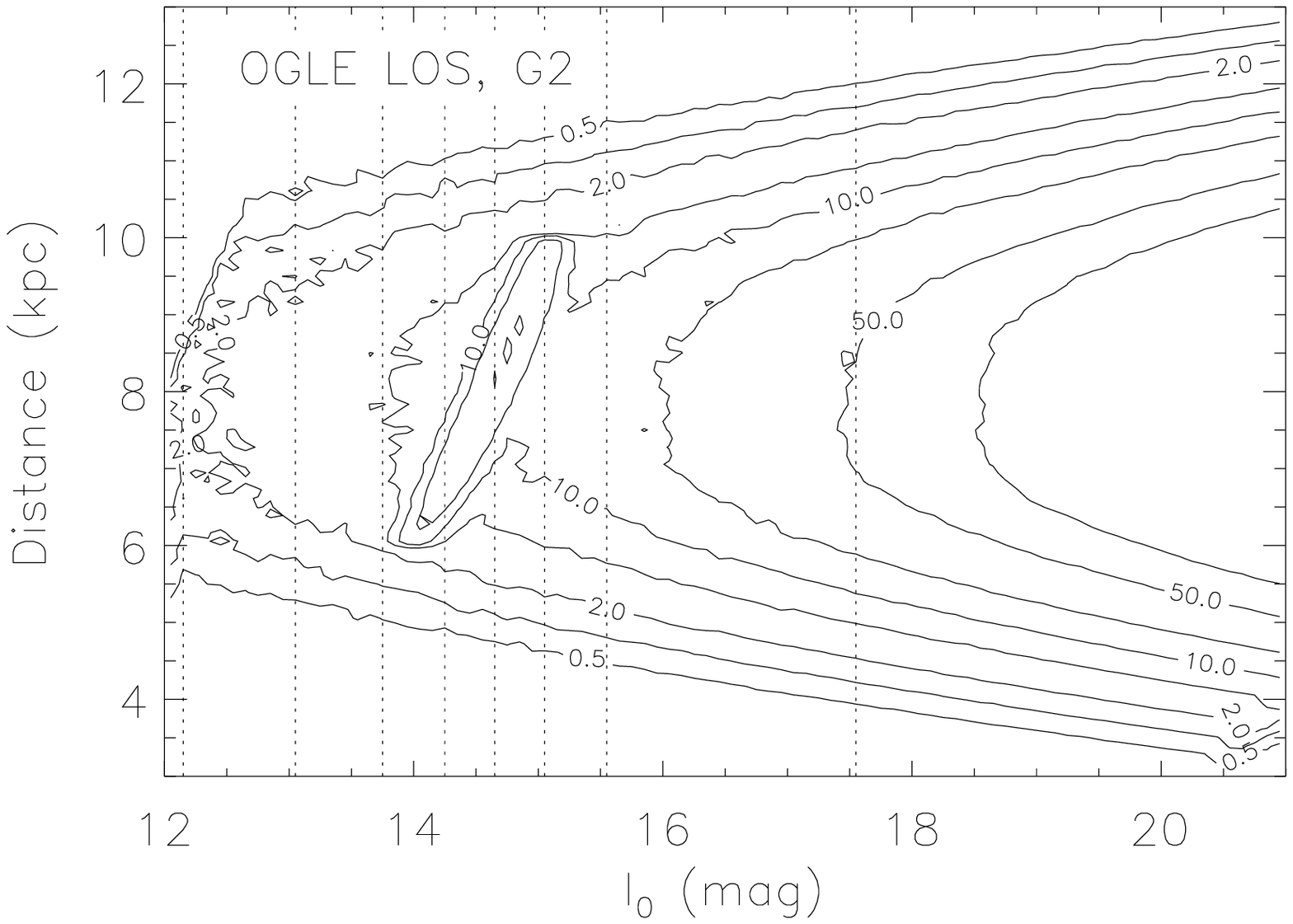}
\includegraphics[width = 7.5cm]{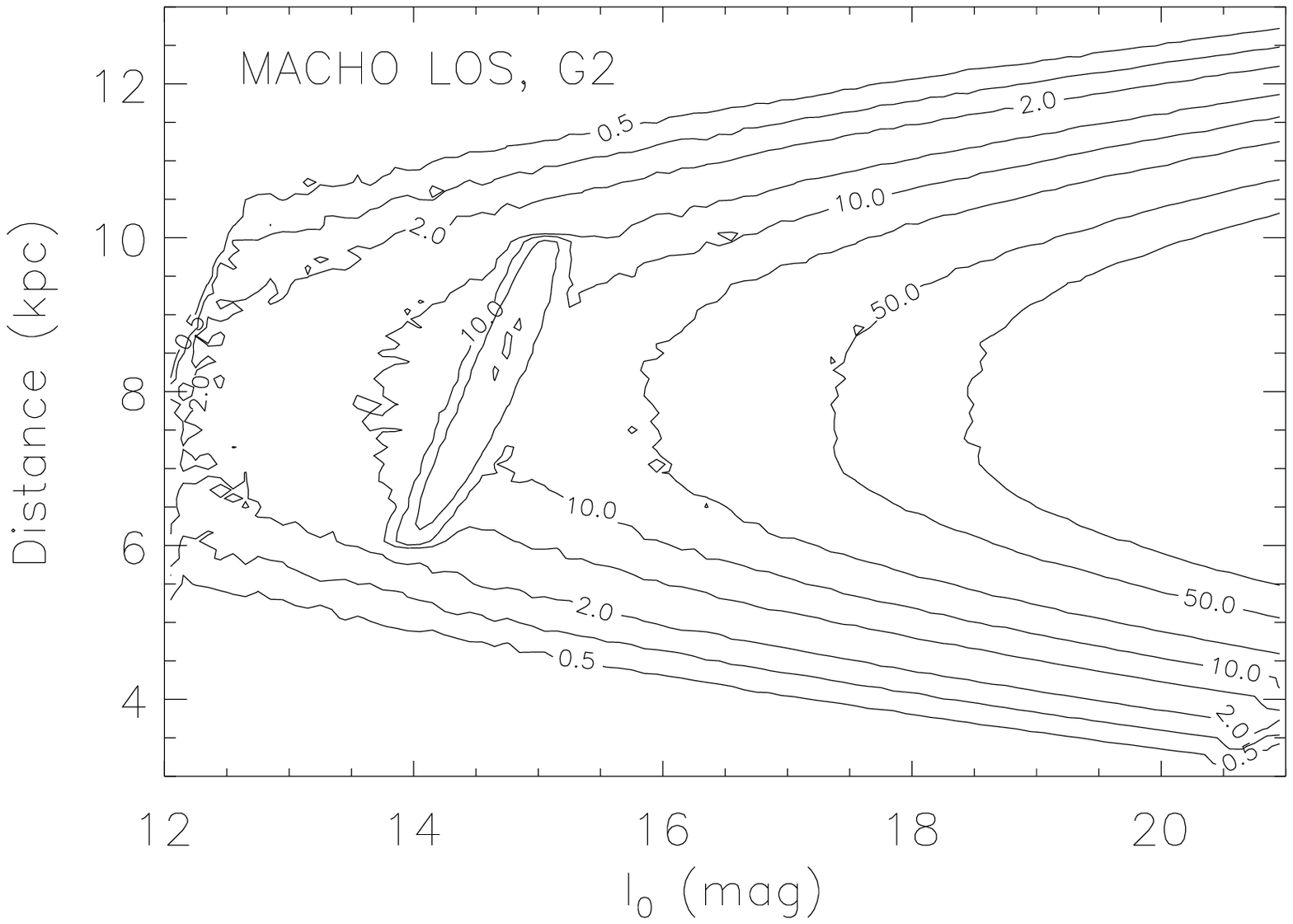}
\includegraphics[width = 7.5cm]{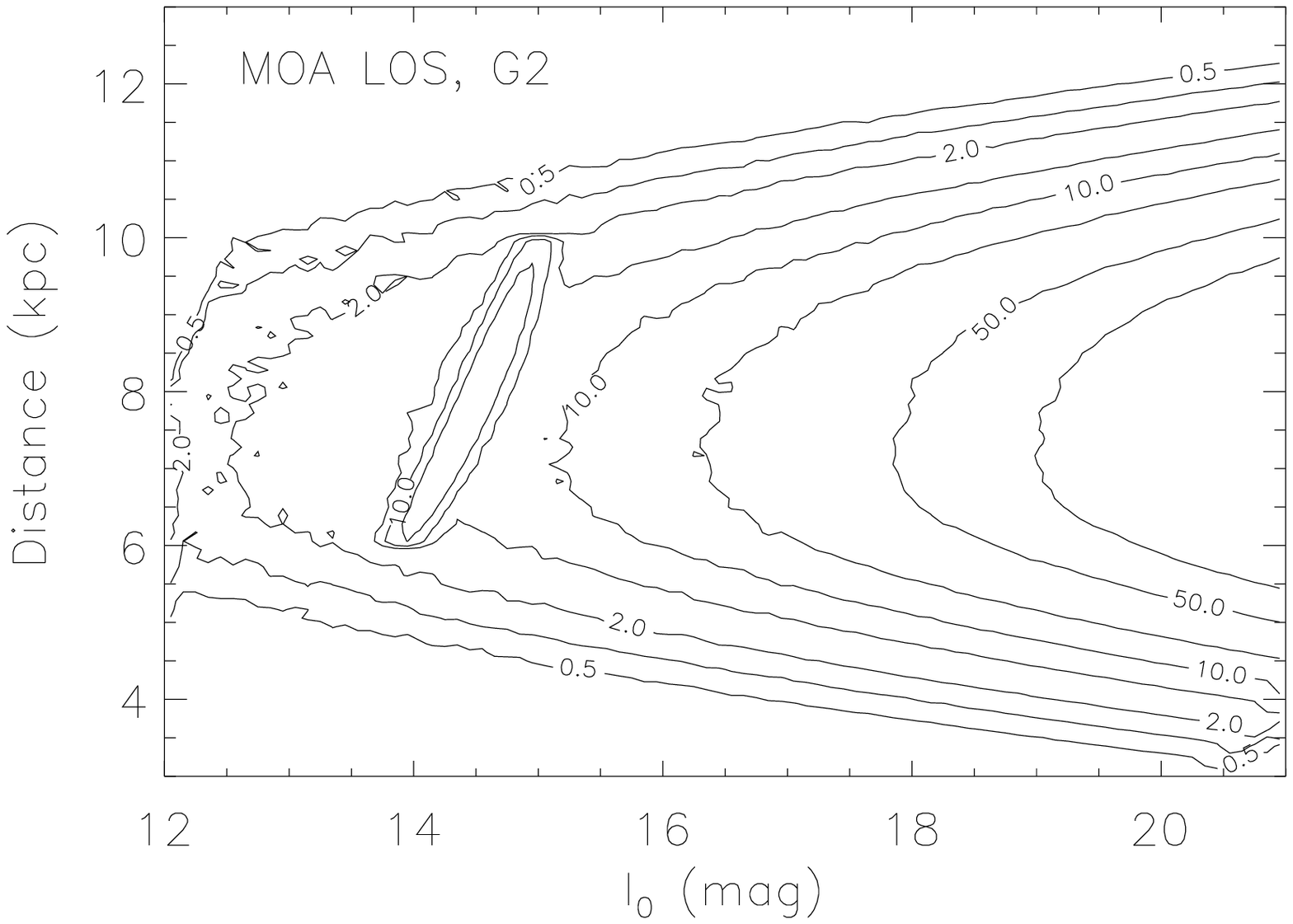}
\caption{(Top, middle, bottom) panel: G2 model source counts as a function of distance 
  and apparent magnitude, for the (OGLE, MACHO, MOA) coordinates. The RCG component is 
  clearly visible (see text). The vertical dotted lines (top panel) correspond to the 
  slices shown in Fig. \ref{fig:magdis_slice_G2} (see text). The normalisation is 
  arbitrary. Contour levels are at 0.5, 1.0, 2.0, 5.0, 10.0, 20.0, 50.0 and 100.0.}
\label{fig:magdis_G2}
\end{figure}

\begin{figure}
\centering
\includegraphics[width = 7.5cm]{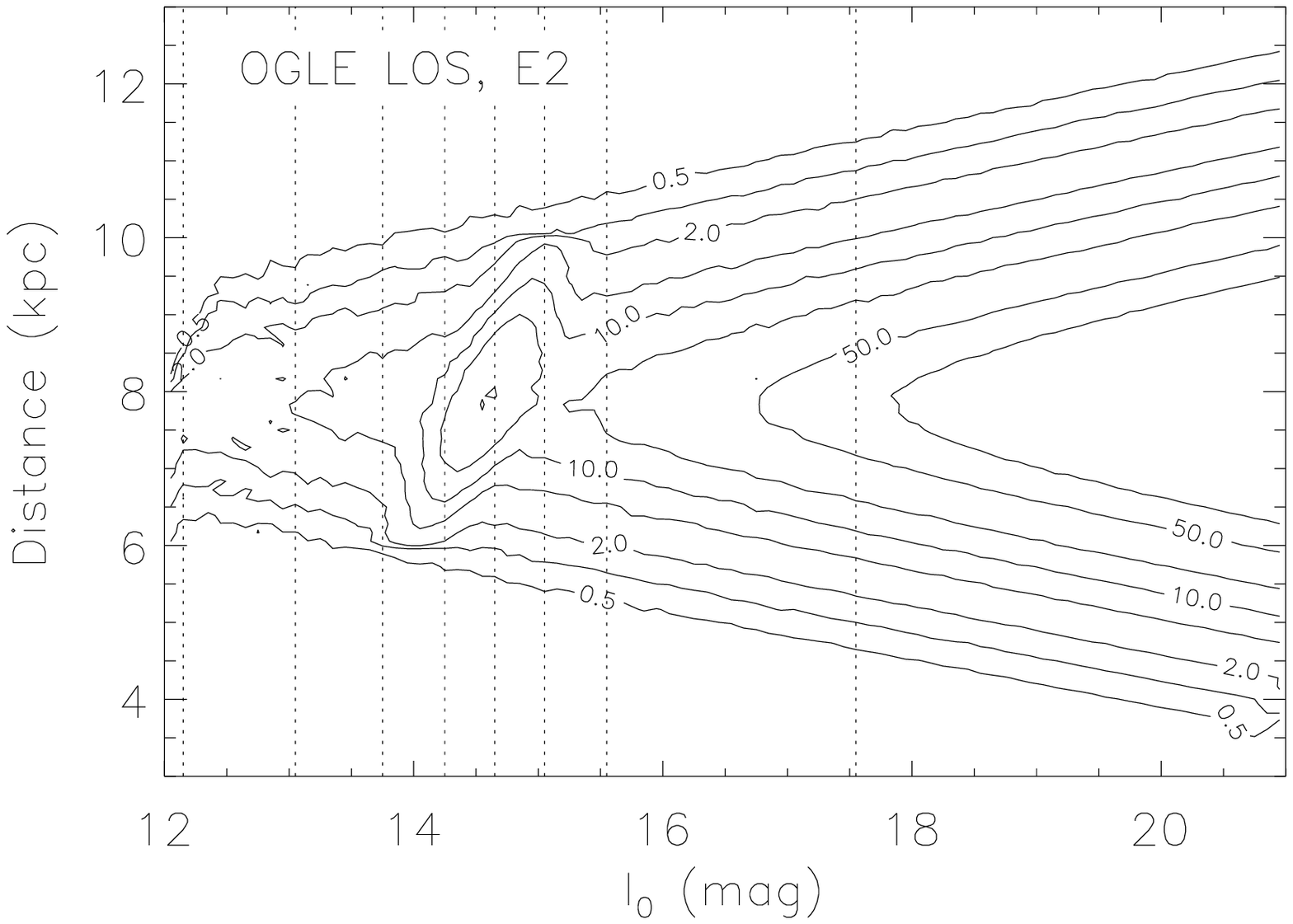}
\includegraphics[width = 7.5cm]{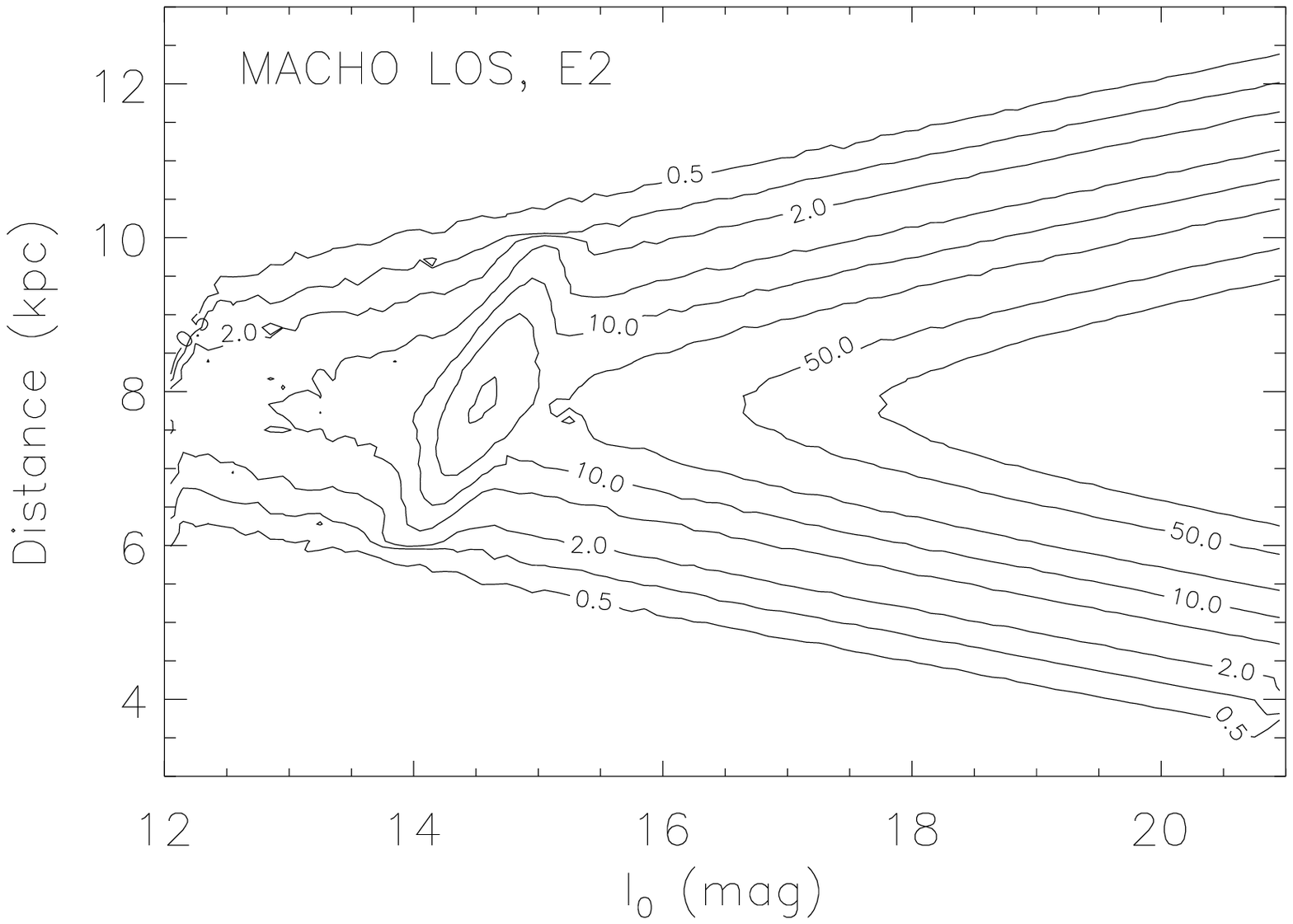}
\includegraphics[width = 7.5cm]{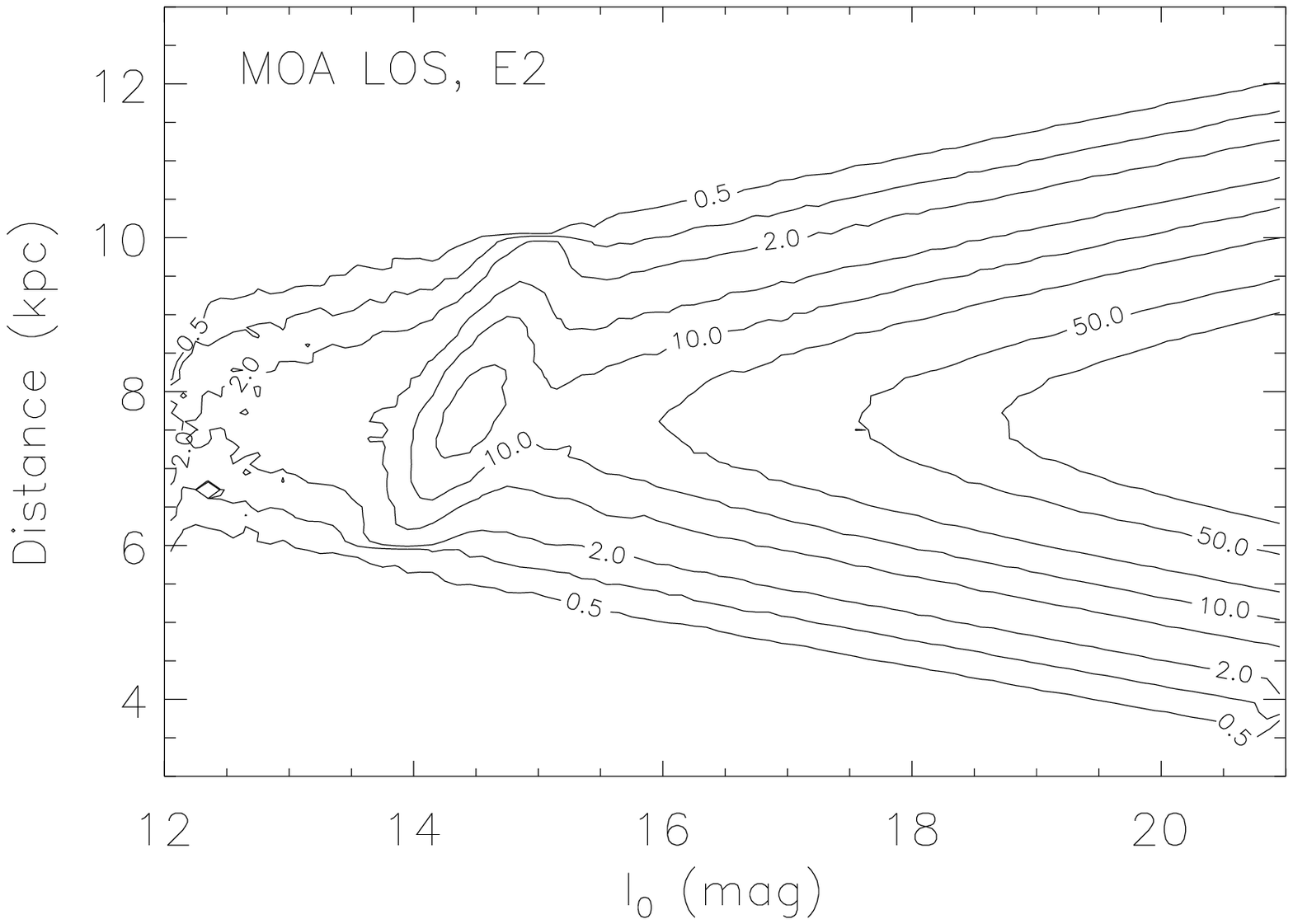}
\caption{Model source counts as a function of distance and apparent magnitude. Same as 
  Fig. \ref{fig:magdis_G2}, but for the E2 model, with the vertical dotted lines 
  corresponding to the slices shown in Fig. \ref{fig:magdis_slice_E2} (see text). The 
  normalisation is arbitrary, but consistent with Fig. \ref{fig:magdis_G2}.}
\label{fig:magdis_E2}
\end{figure}

\begin{figure}
\centering
\includegraphics[width = 7.5cm]{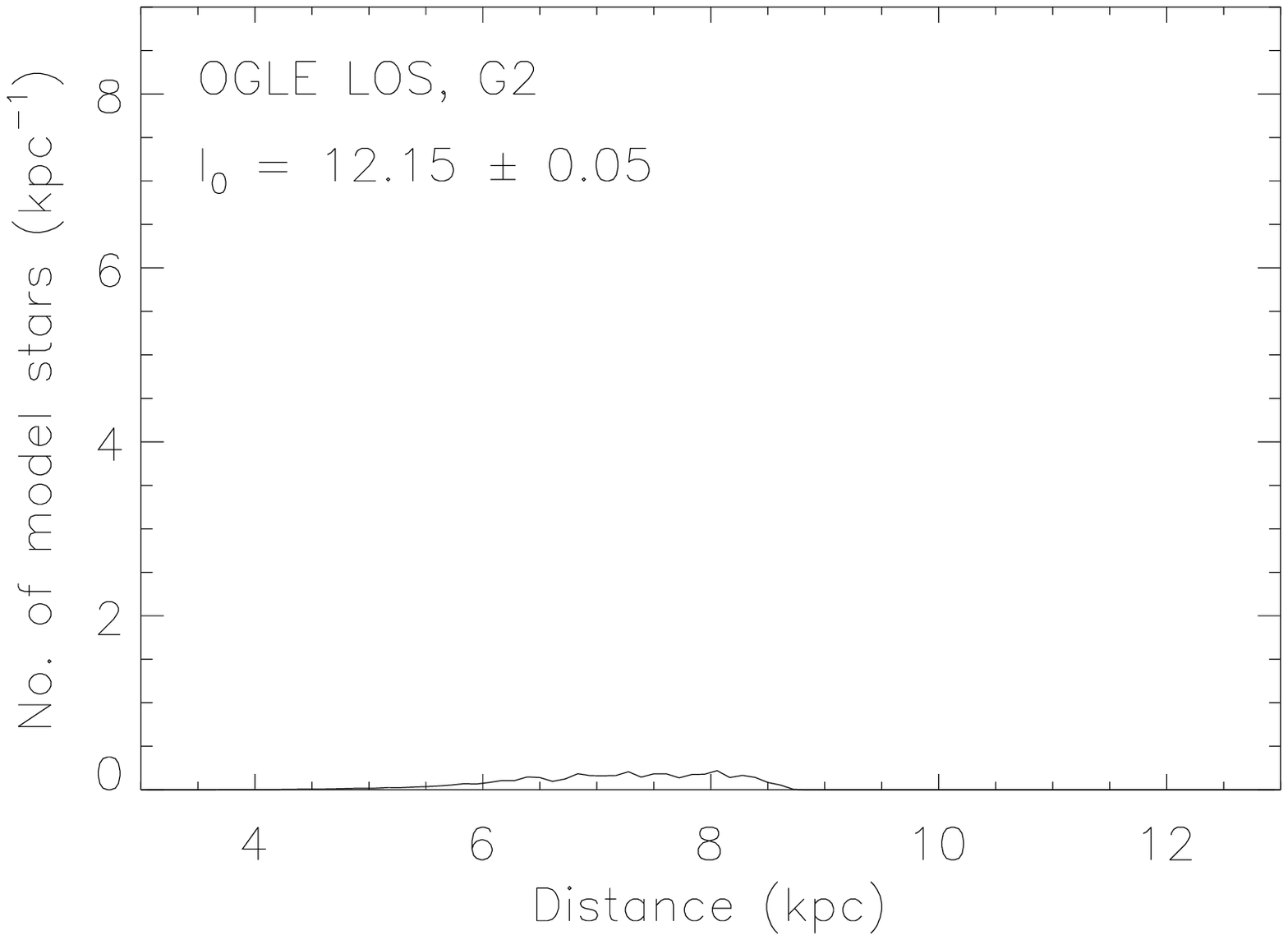}
\includegraphics[width = 7.5cm]{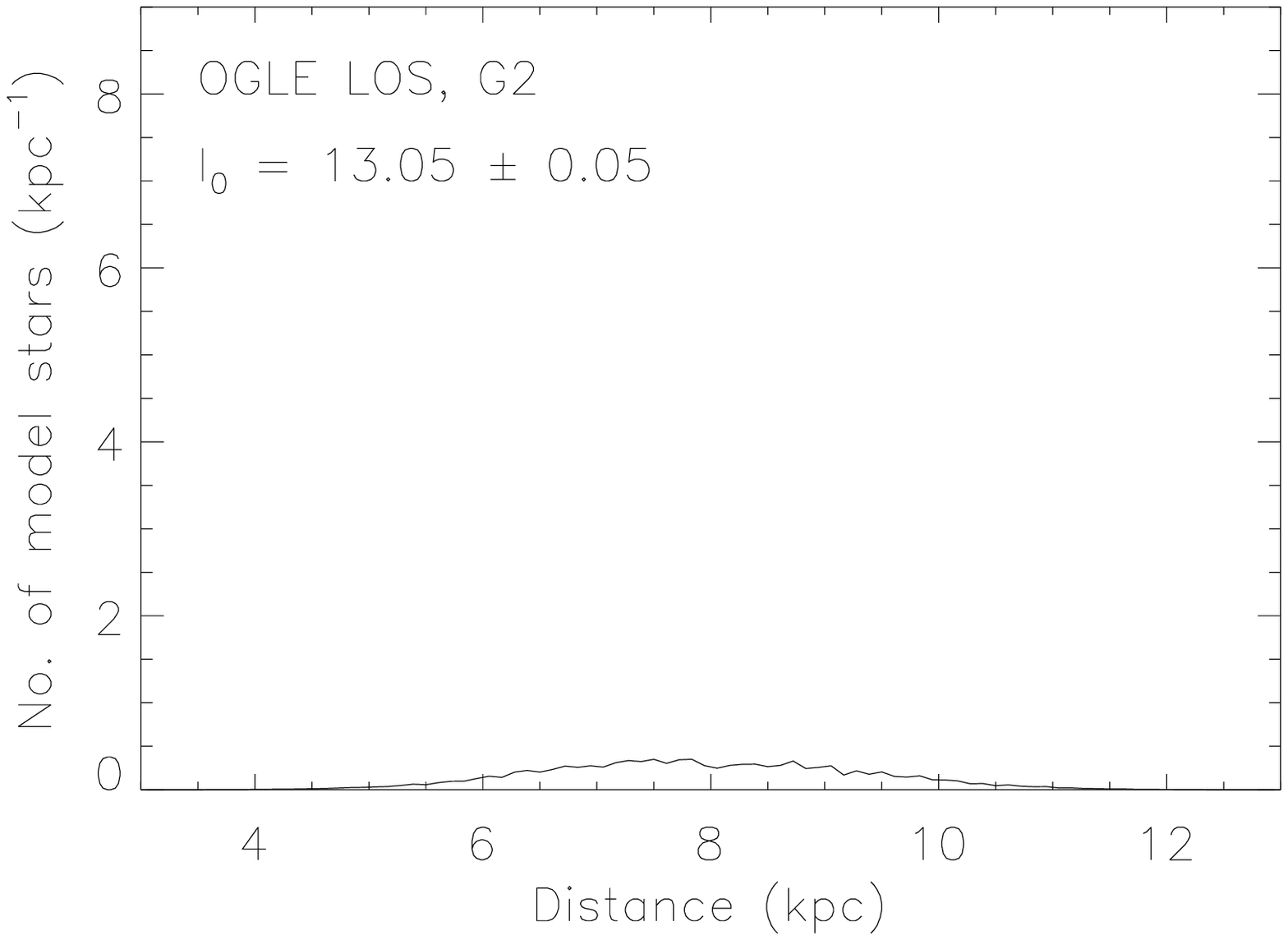}
\includegraphics[width = 7.5cm]{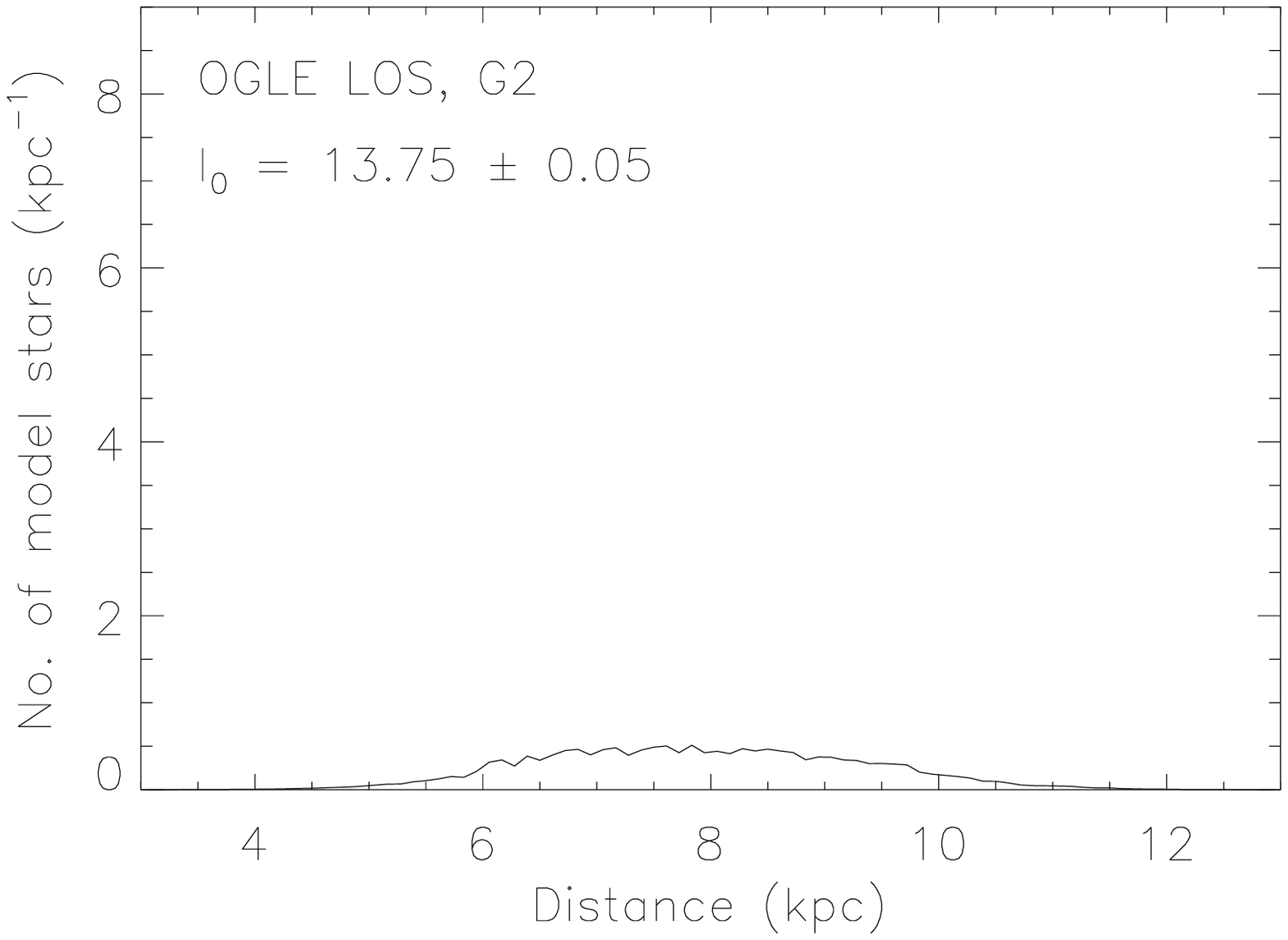}
\includegraphics[width = 7.5cm]{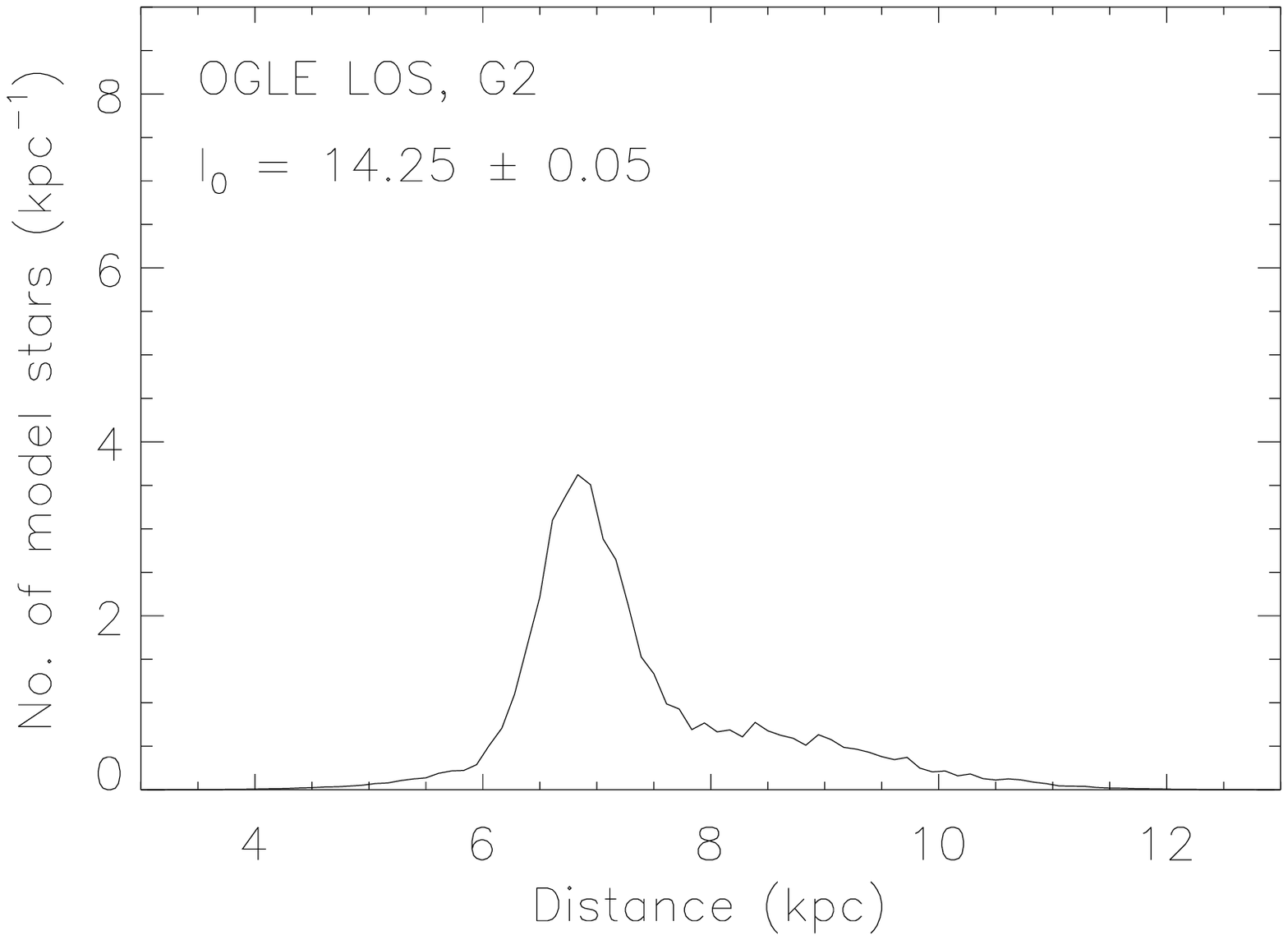}
\includegraphics[width = 7.5cm]{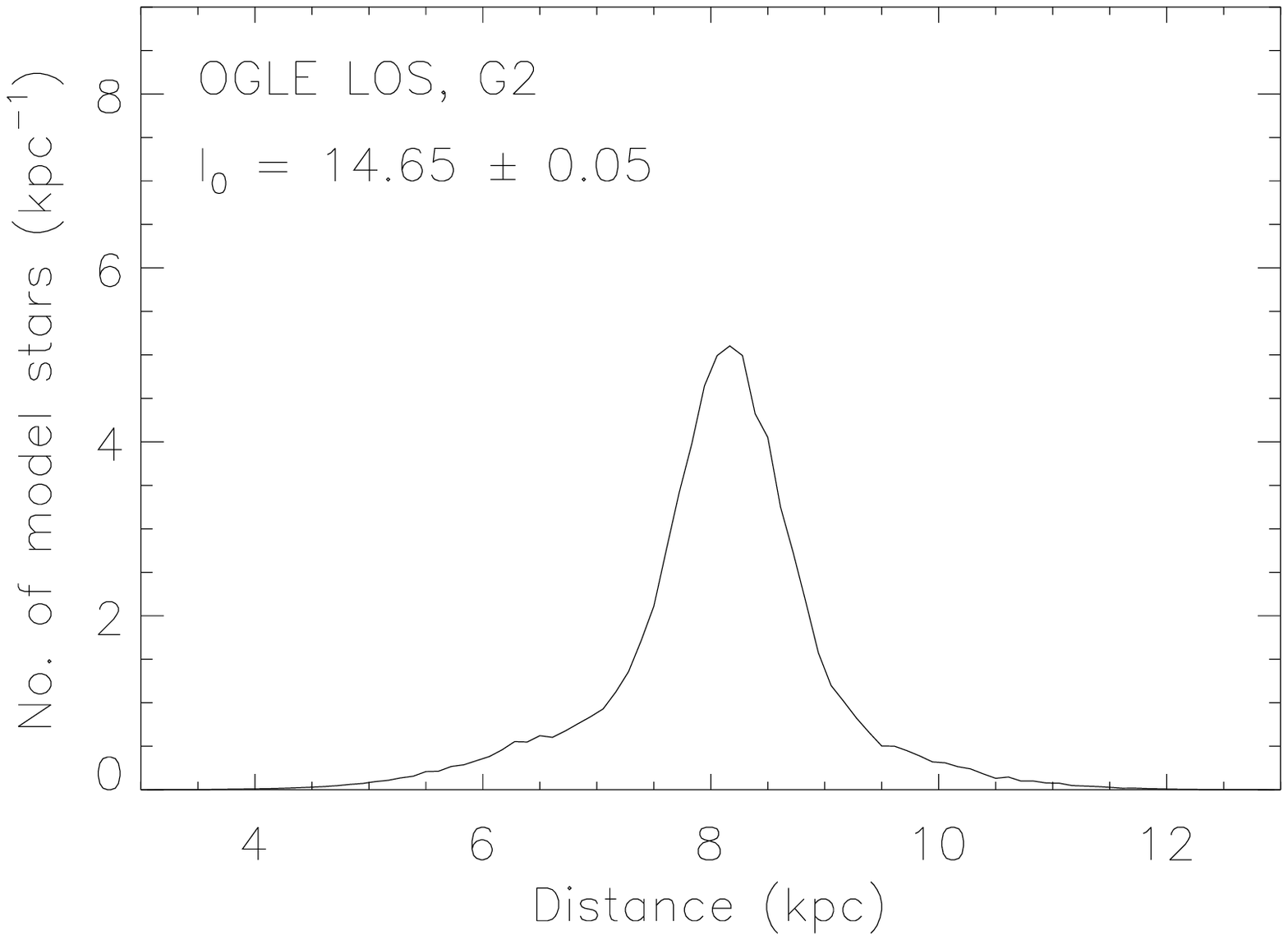}
\includegraphics[width = 7.5cm]{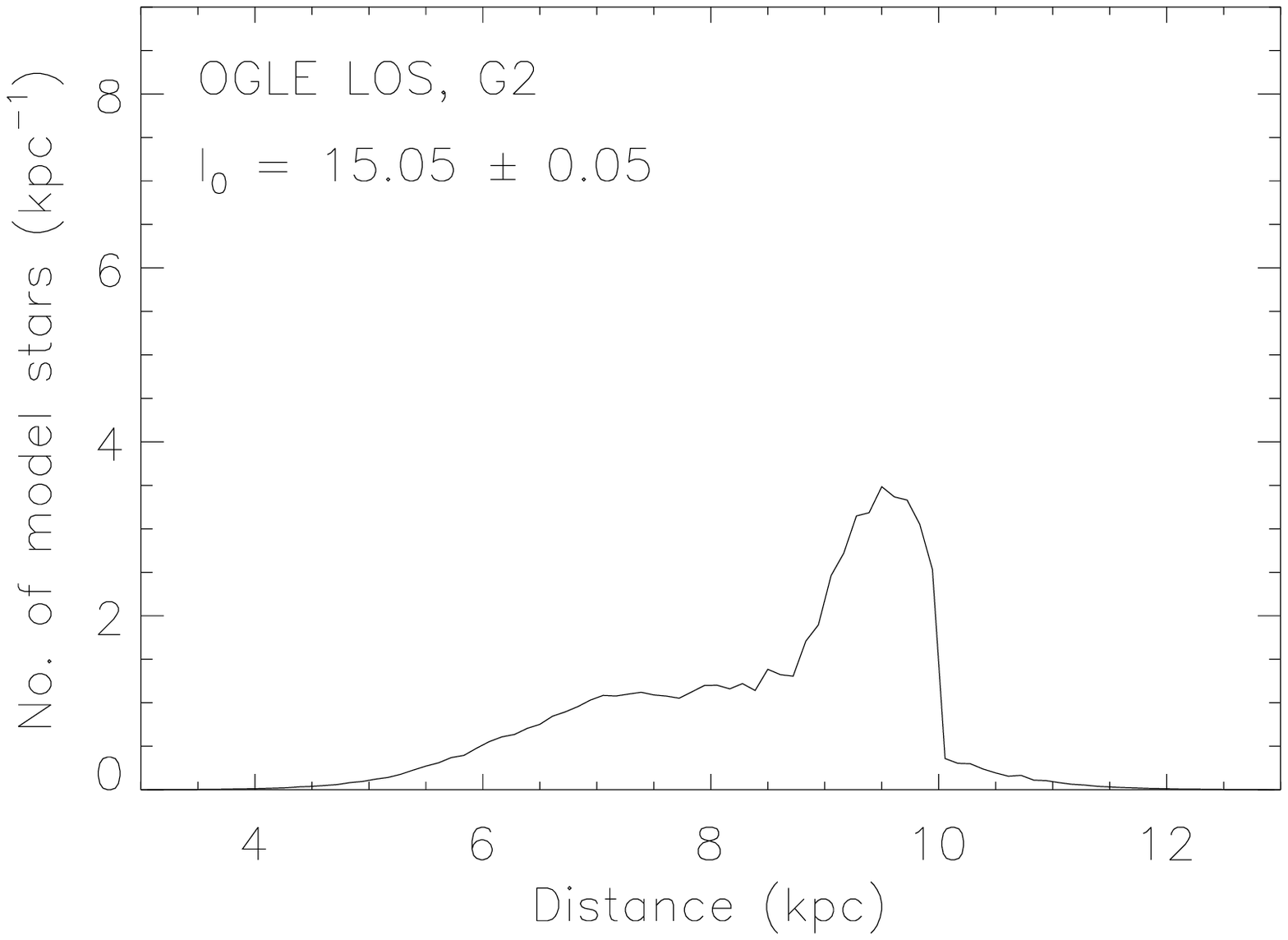}
\includegraphics[width = 7.5cm]{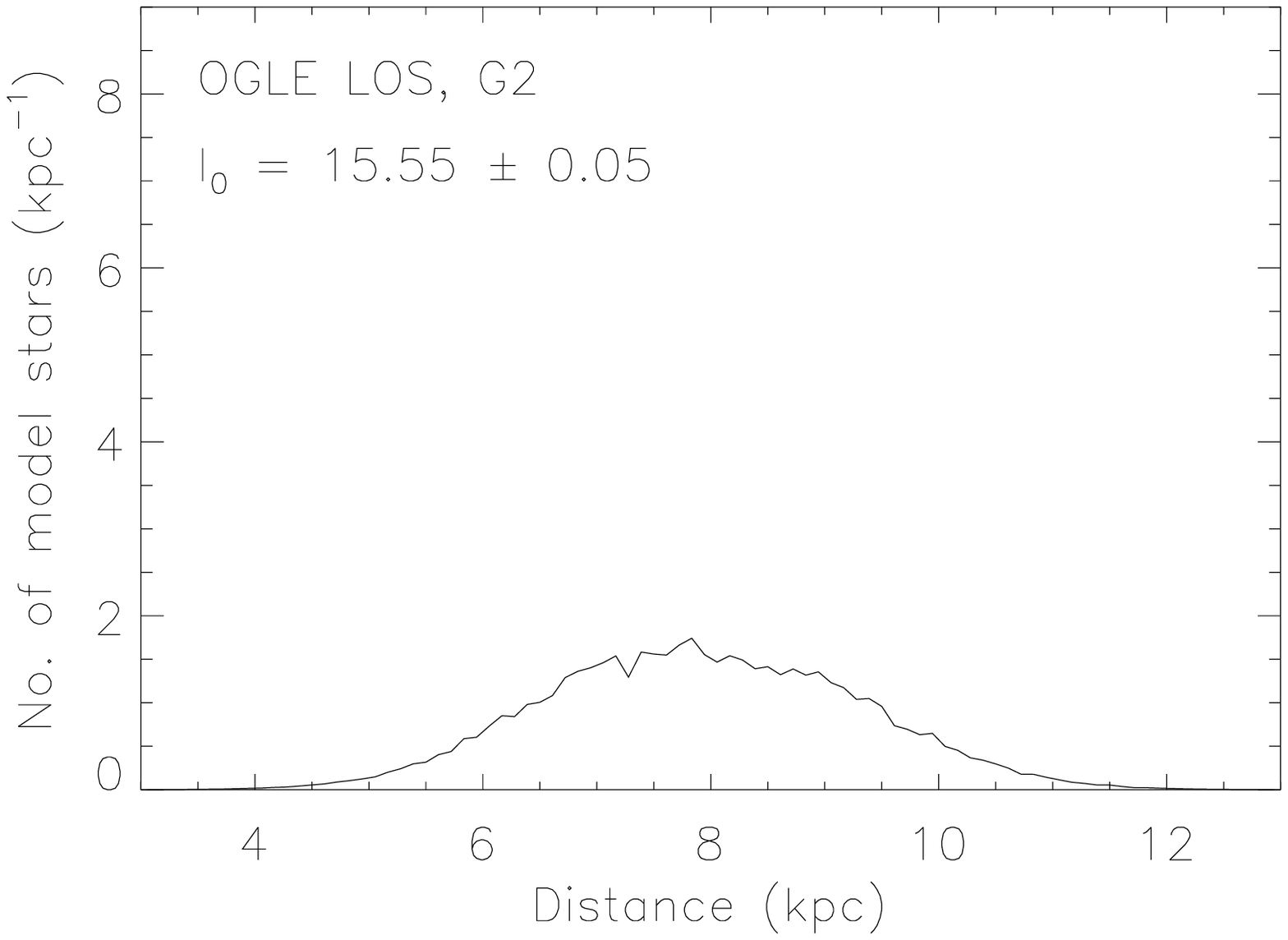}
\includegraphics[width = 7.5cm]{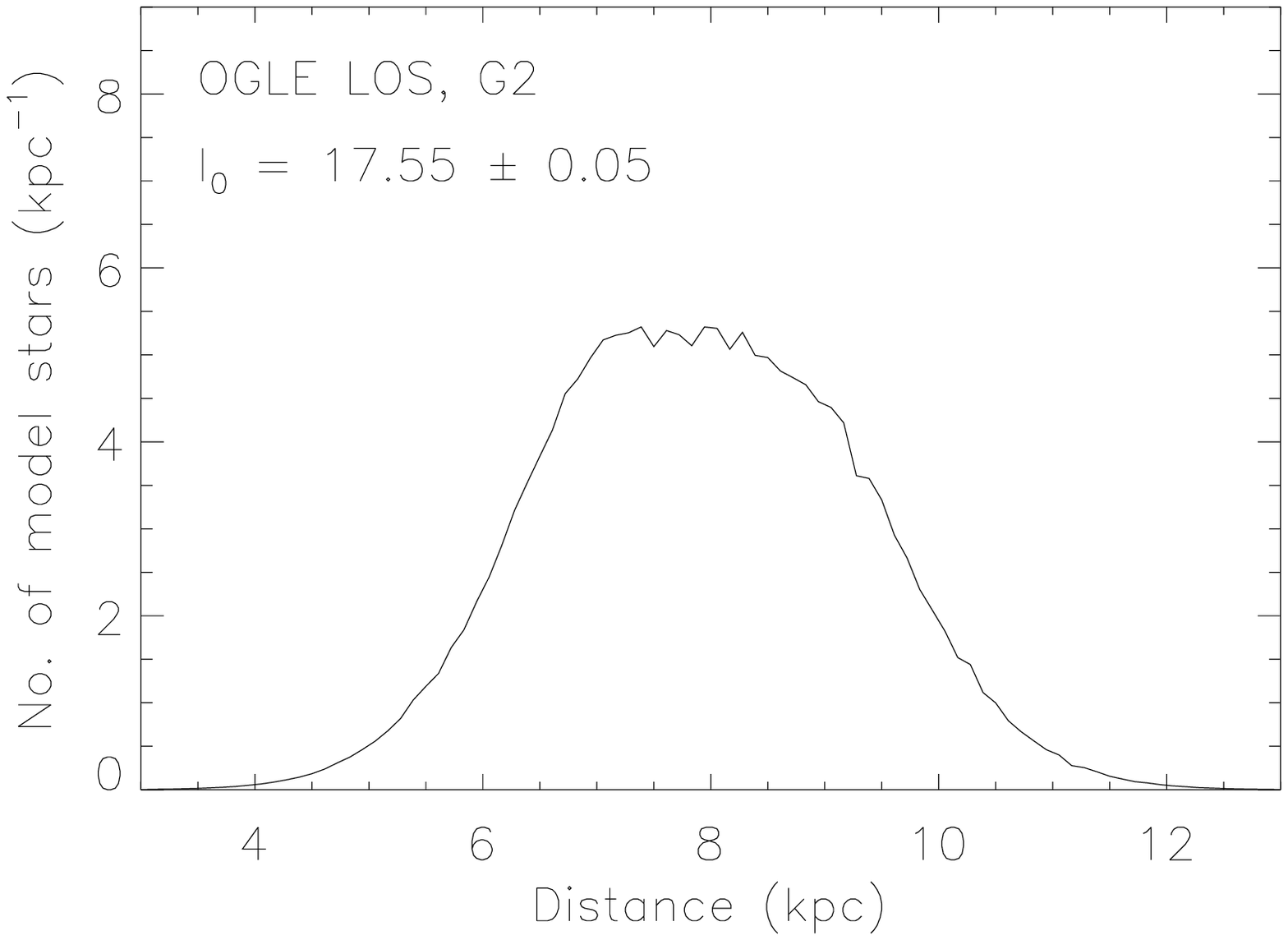}
\caption{Model source counts as a function of distance (OGLE LOS, G2 model), for 
  selected apparent magnitudes $I_0$ (with $I_0$ bin widths of 0.1 mag). These 
  magnitudes are indicated in the panels, and correspond to the slices indicated in 
  Fig. \ref{fig:magdis_G2} (top panel).}
\label{fig:magdis_slice_G2}
\end{figure}

\begin{figure}
\centering
\includegraphics[width = 7.5cm]{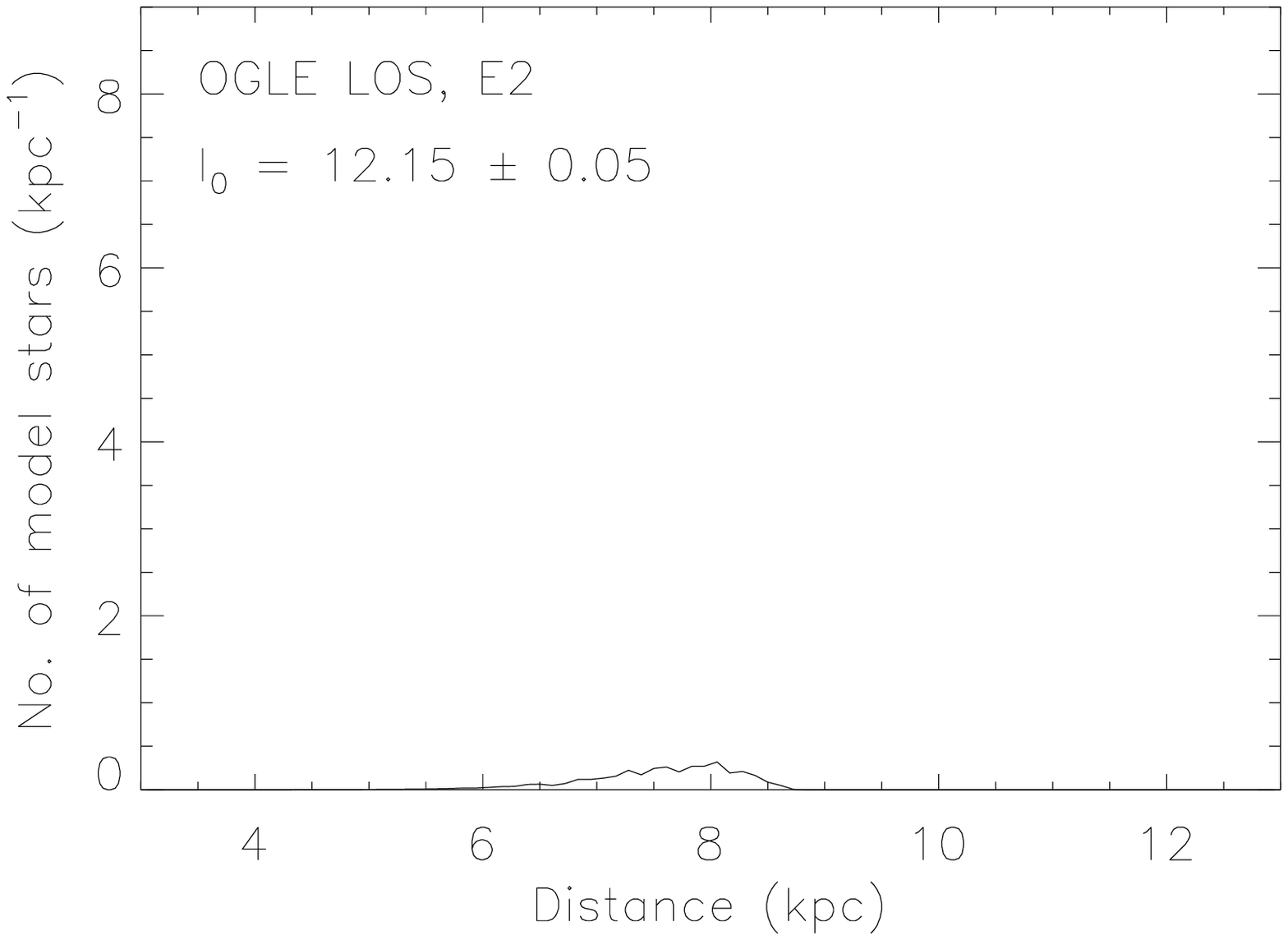}
\includegraphics[width = 7.5cm]{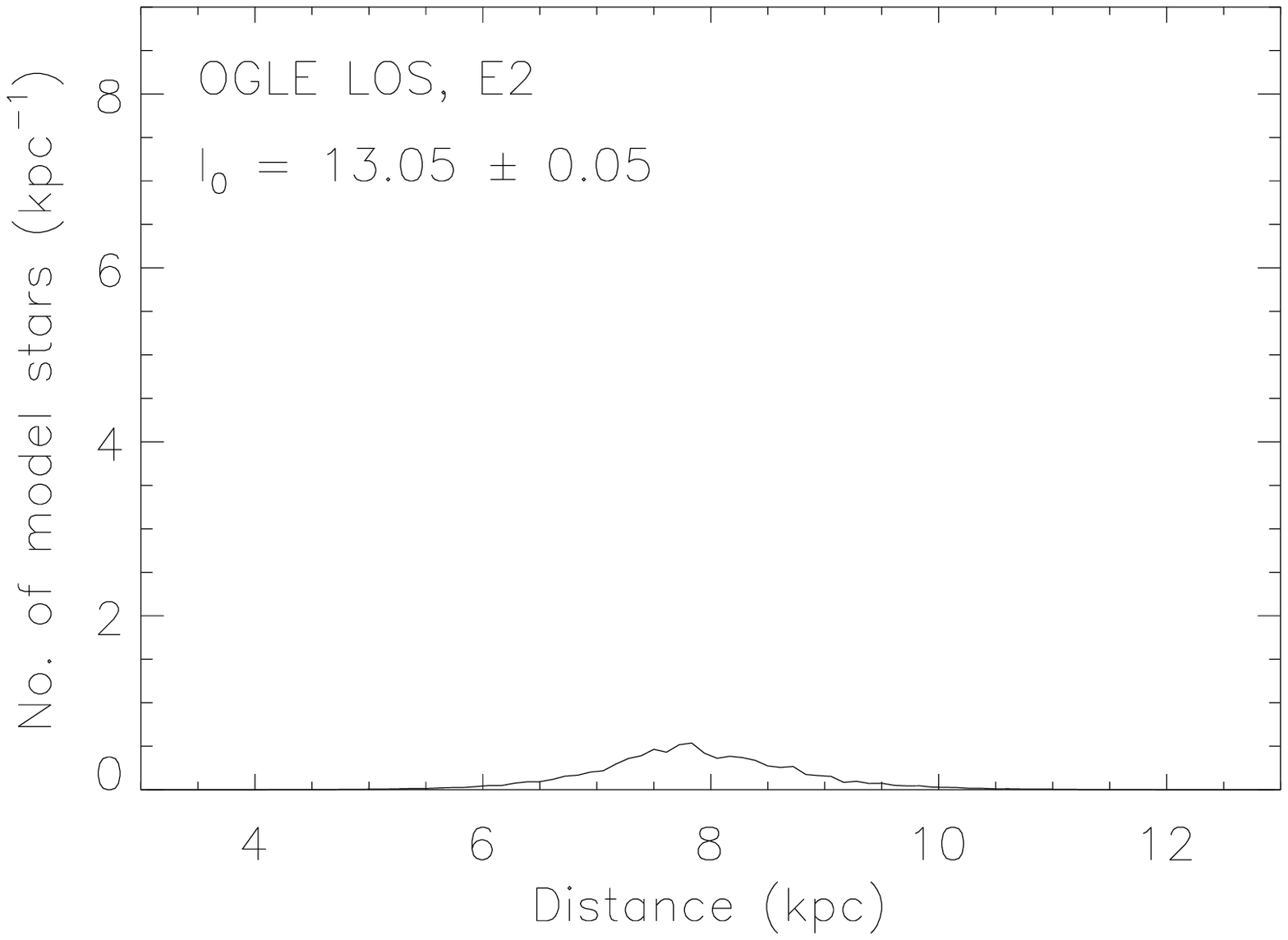}
\includegraphics[width = 7.5cm]{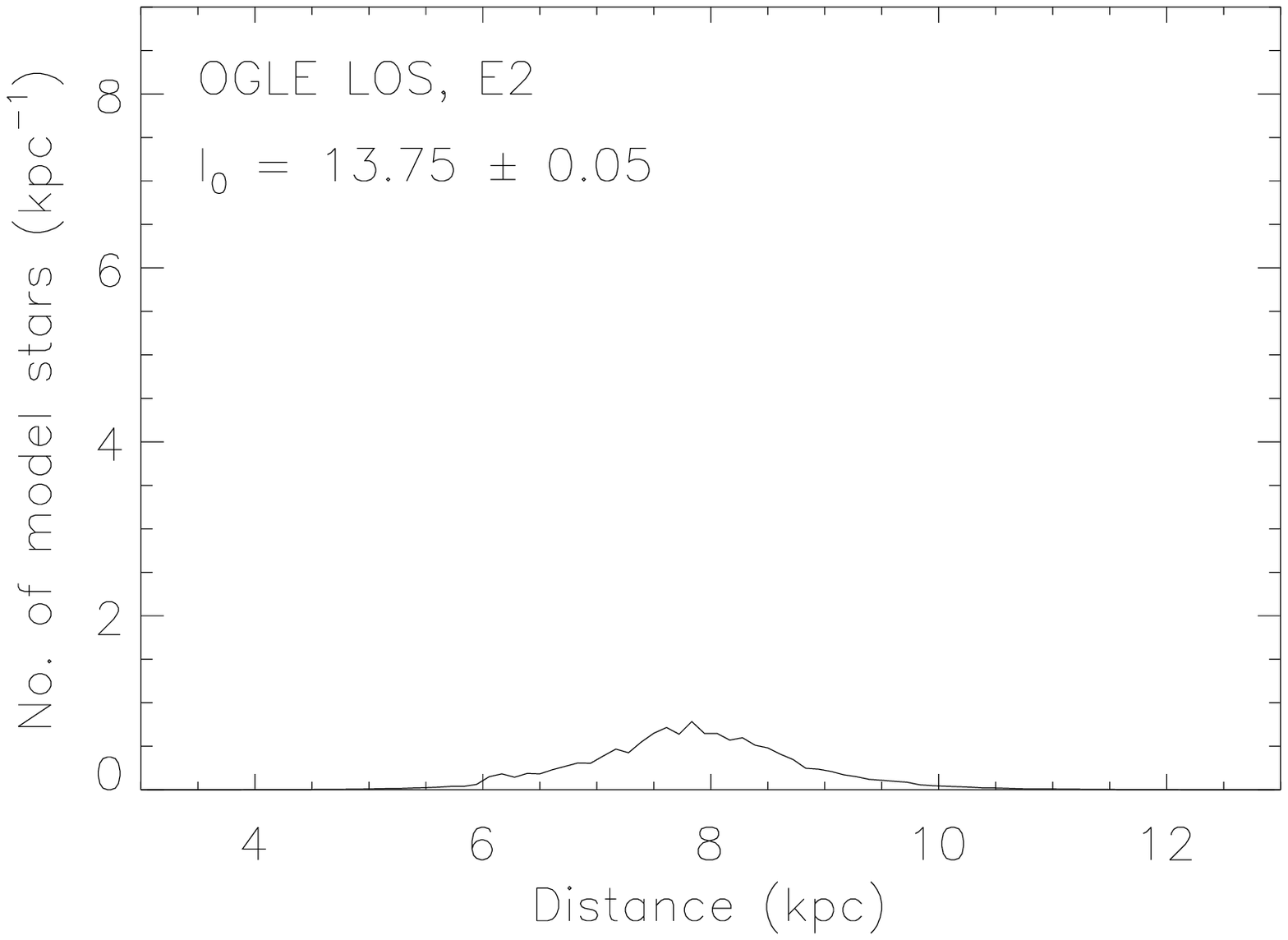}
\includegraphics[width = 7.5cm]{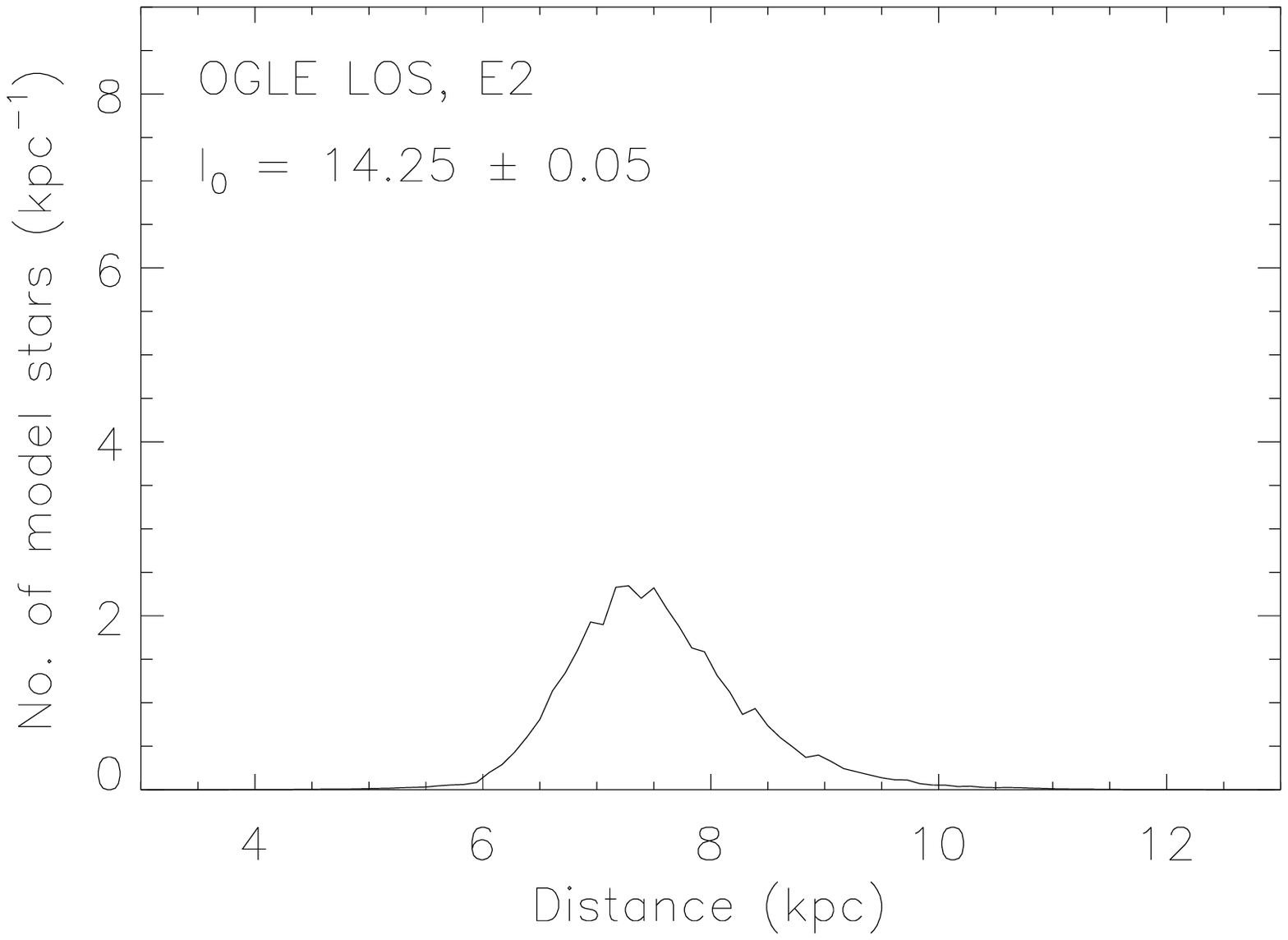}
\includegraphics[width = 7.5cm]{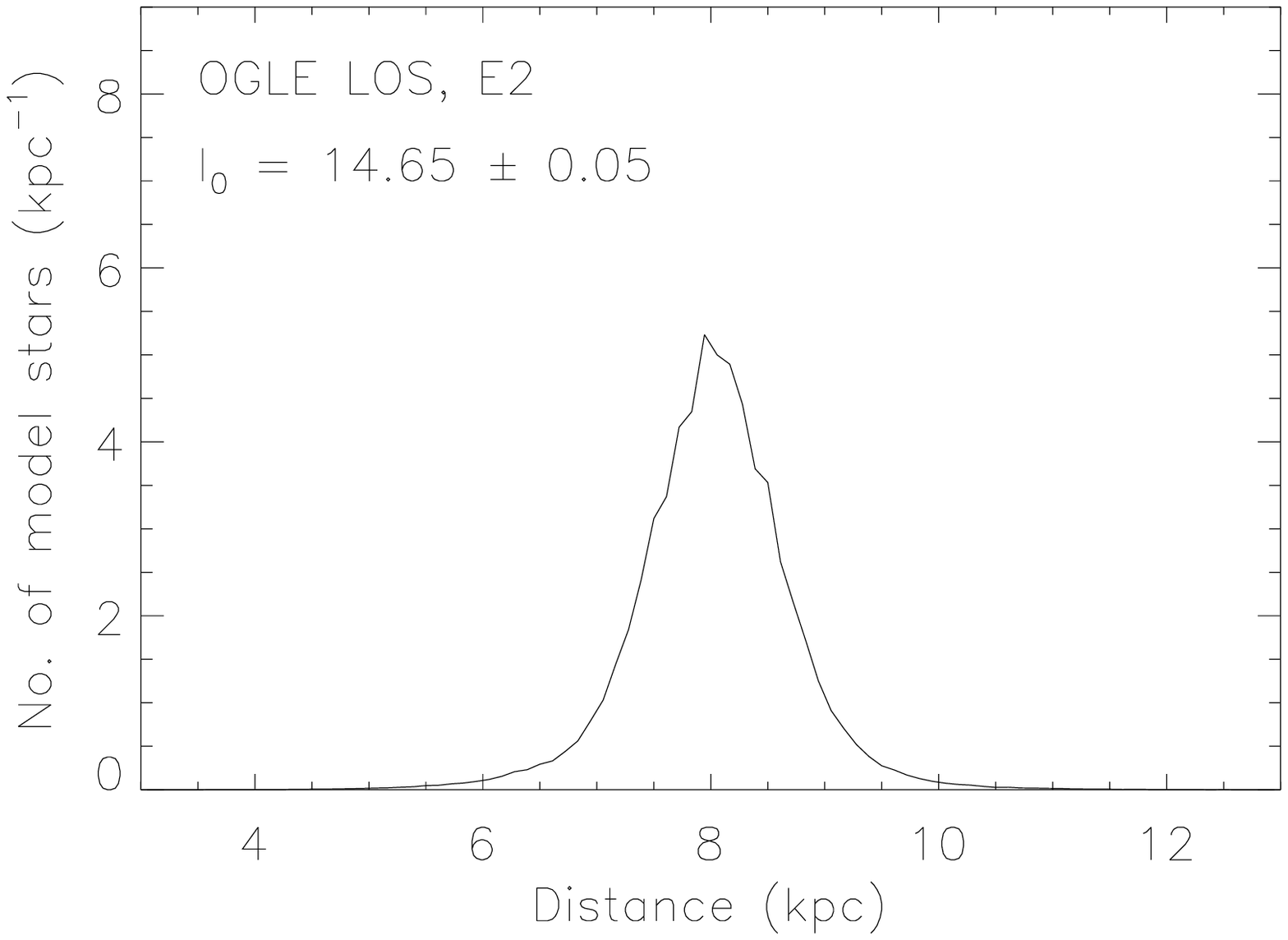}
\includegraphics[width = 7.5cm]{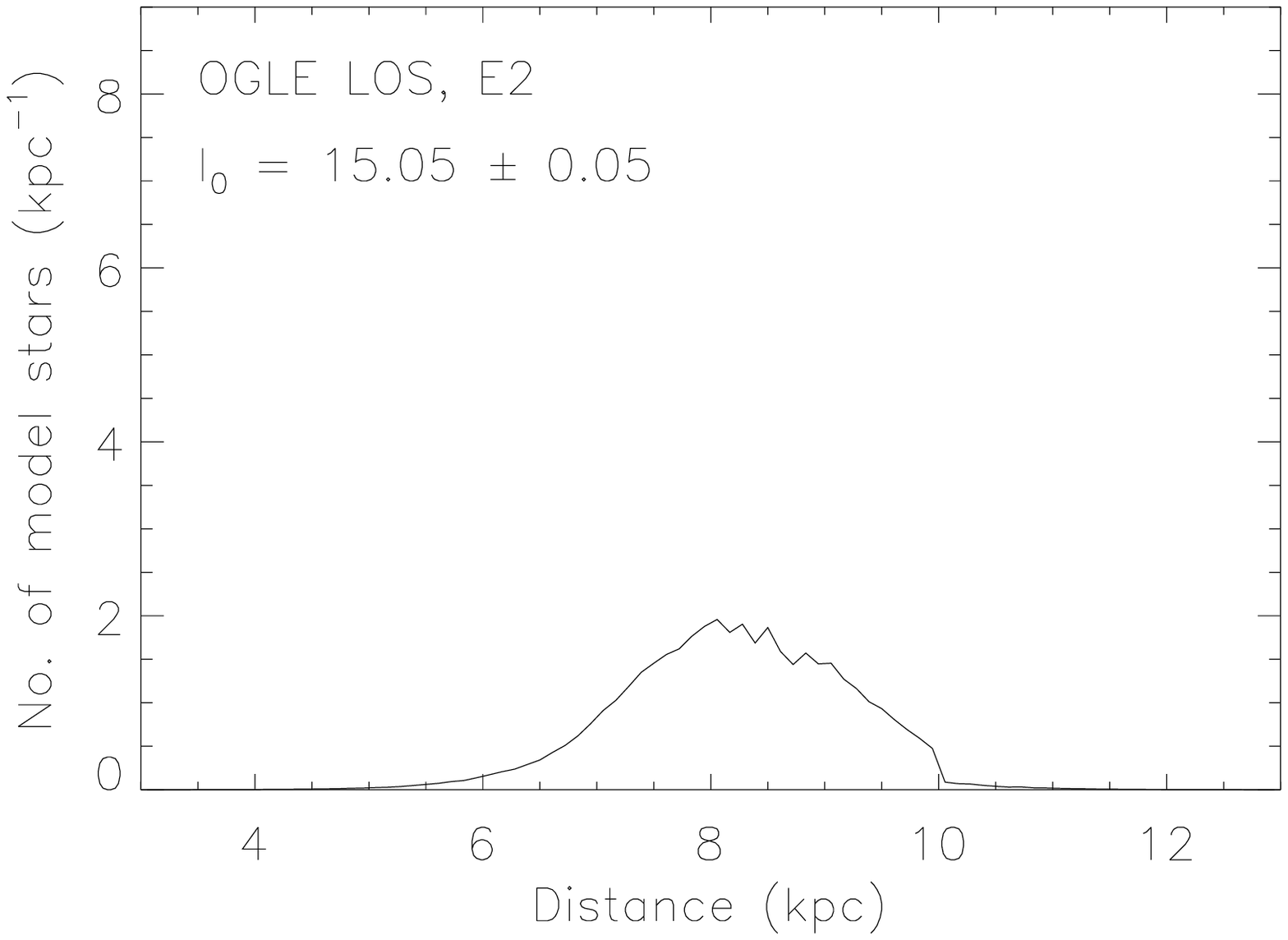}
\includegraphics[width = 7.5cm]{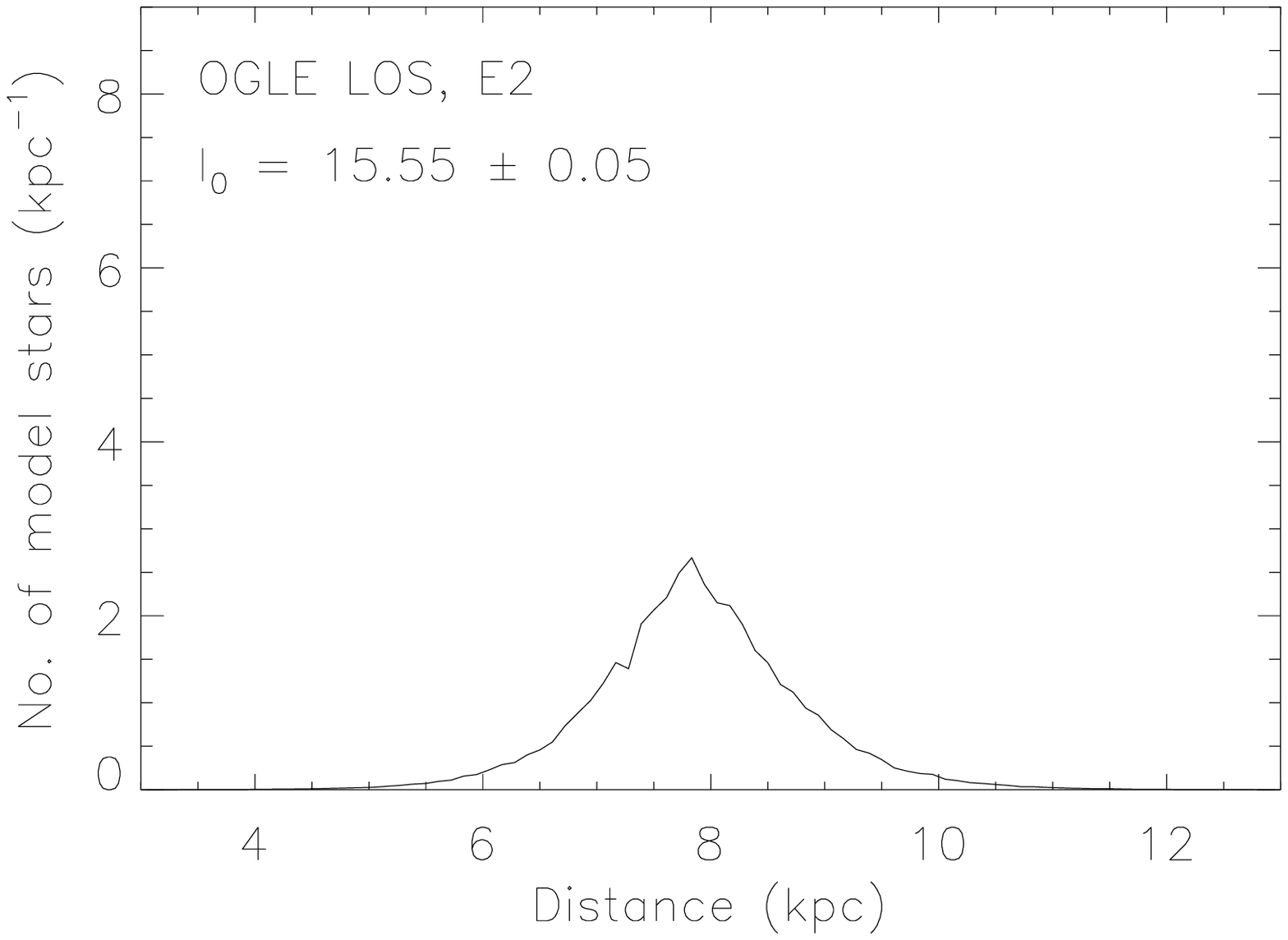}
\includegraphics[width = 7.5cm]{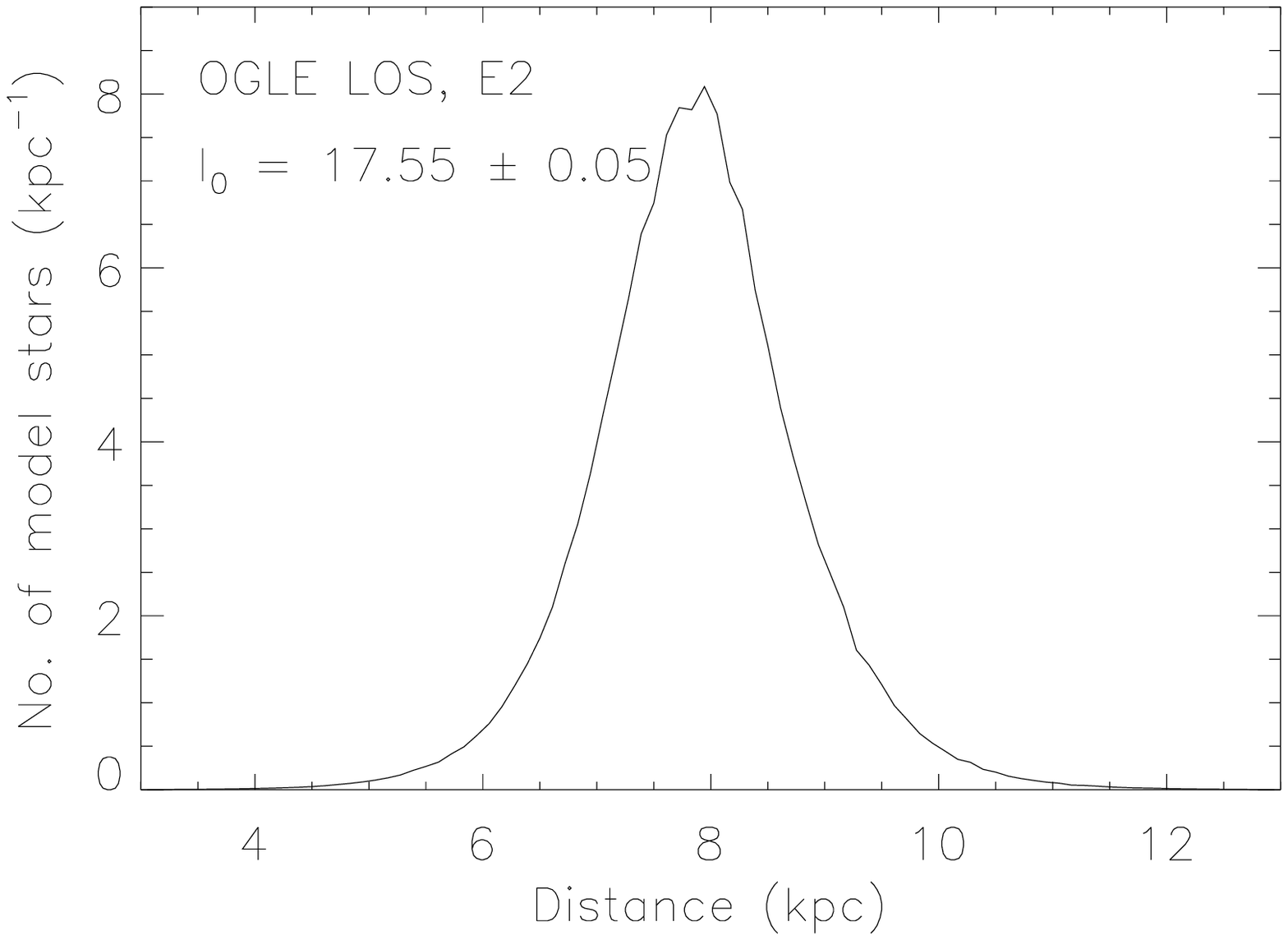}
\caption{Model source counts as a function of distance. Same as Fig. 
  \ref{fig:magdis_slice_G2}, but for the E2 model, with the selected magnitudes 
  corresponding to the slices indicated in Fig. \ref{fig:magdis_E2} (top panel).}
\label{fig:magdis_slice_E2}
\end{figure}

\begin{figure}
\centering
\includegraphics[width = 7.5cm]{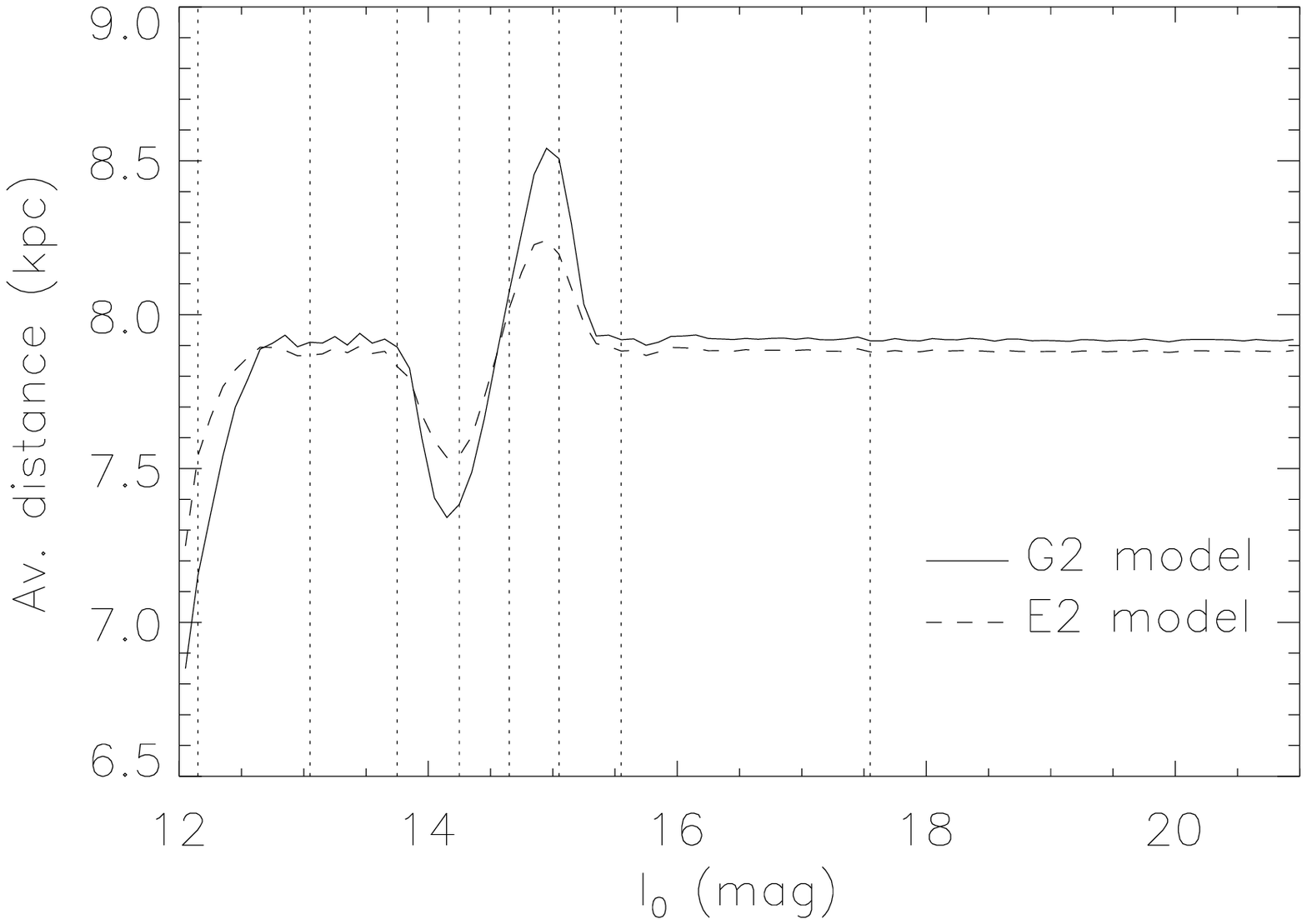}
\includegraphics[width = 7.5cm]{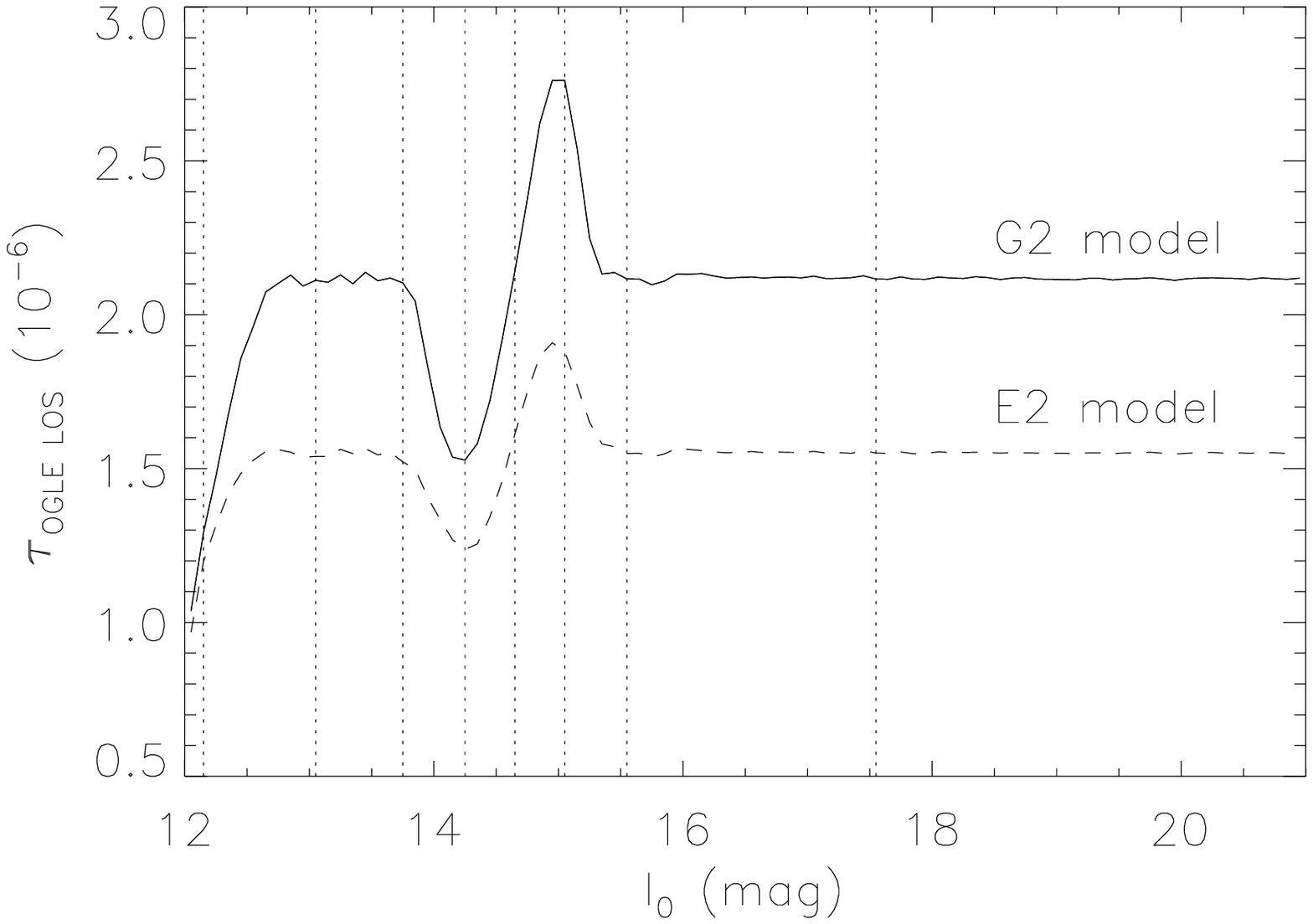}
\caption{Top panel: average distance of model OGLE LOS sources as a function of 
  magnitude. Bottom panel: $\tauoglelos$ as a function of magnitude (same as top 
  panel of Fig. \ref{fig:magtau}). The vertical dotted lines correspond to the 
  slice magnitudes indicated in Figs. \ref{fig:magdis_G2}--\ref{fig:magdis_slice_E2}.}
\label{fig:av_dis}
\end{figure}

This strong correlation displayed by the RCGs is illustrated as follows. In Figs. 
\ref{fig:magdis_G2} and \ref{fig:magdis_E2}, contours are plotted of source counts as a 
function of distance and apparent magnitude along each LOS, for the G2 and E2 models, 
respectively. The two components of the source population are clearly distinguishable. 
For the G2 model, most of the RCGs appear as a narrow diagonal line in the range $14.0 
\lesssim I_0 \lesssim 15.0$. For E2, this region is wider (for a given distance) and 
shallower. These differences in shape are primarily due to the different bar angles of 
the G2 and E2 models ($13.4^\circ$ and $41.3^\circ$, respectively). The red giants and 
other stars form a smoother background, with a steep increase in numbers, and a 
broadening in distance, as $I_0$ increases. In the top panels of Figs. 
\ref{fig:magdis_G2} and \ref{fig:magdis_E2}, the vertical dotted lines indicate slices 
of this distribution -- for the OGLE LOS -- that are shown in Figs. 
\ref{fig:magdis_slice_G2} and \ref{fig:magdis_slice_E2}, respectively. Finally, the top 
panel of Fig. \ref{fig:av_dis} gives the average distance of model OGLE LOS sources as 
a function of $I_0$. The slice magnitudes are indicated, and the bottom panel shows how 
they intersect the $\tauoglelos$ trend from Fig. \ref{fig:magtau}. Note that the trends 
of average source distance and optical depth with $I_0$ are almost identical, as would 
be expected for the reasons given above.

The expected oscillation in $\tau$ caused by the RCGs is clearly significant. For 
example, Fig. \ref{fig:magtau} shows that the $\tau$ amplitude along the OGLE LOS is 
$\sim$$1.2 \times 10^{-6}$. This is a deviation of $\sim$$\pm 30$ per cent from the 
approximately constant optical depth at fainter magnitudes (hereafter 
\emph{$\tauflat$}), where far fewer RCGS are seen: $\tauflat \approx 2.1\times 
10^{-6}$. For comparison, OGLE's measured value of $2.55^{+0.57}_{-0.46} \times 
10^{-6}$ has an uncertainty of only $\sim$$\pm 20$ per cent, so the predicted 
oscillation ought to be detectable if enough sources are observed at the correct 
magnitudes.

\onecolumn

\twocolumn

\begin{table}
\centering
\begin{tabular}{lccc}\hline

  & ($l, b$) ($^\circ$) & $\theta_{\rm bar, G2}$ ($^\circ$) & $\theta_{\rm bar, E2}$ ($^\circ$) \\ \hline

  OGLE  & (1.16, $-2.75$) & $14.8$ & $19.1$ \\
  MACHO & (1.50, $-2.68$) & $24.6$ & $31.0$ \\ \hline

\end{tabular}
\caption{$1 \sigma$ upper limits on $\thetabar$, from combining the expected optical 
  depths with those measured by OGLE and MACHO (see text). Values are shown for the 
  G2 and E2 bar models.}
\label{tab:thetabar_limits}
\end{table}

\begin{figure}
\centering
\includegraphics[width = 7.5cm]{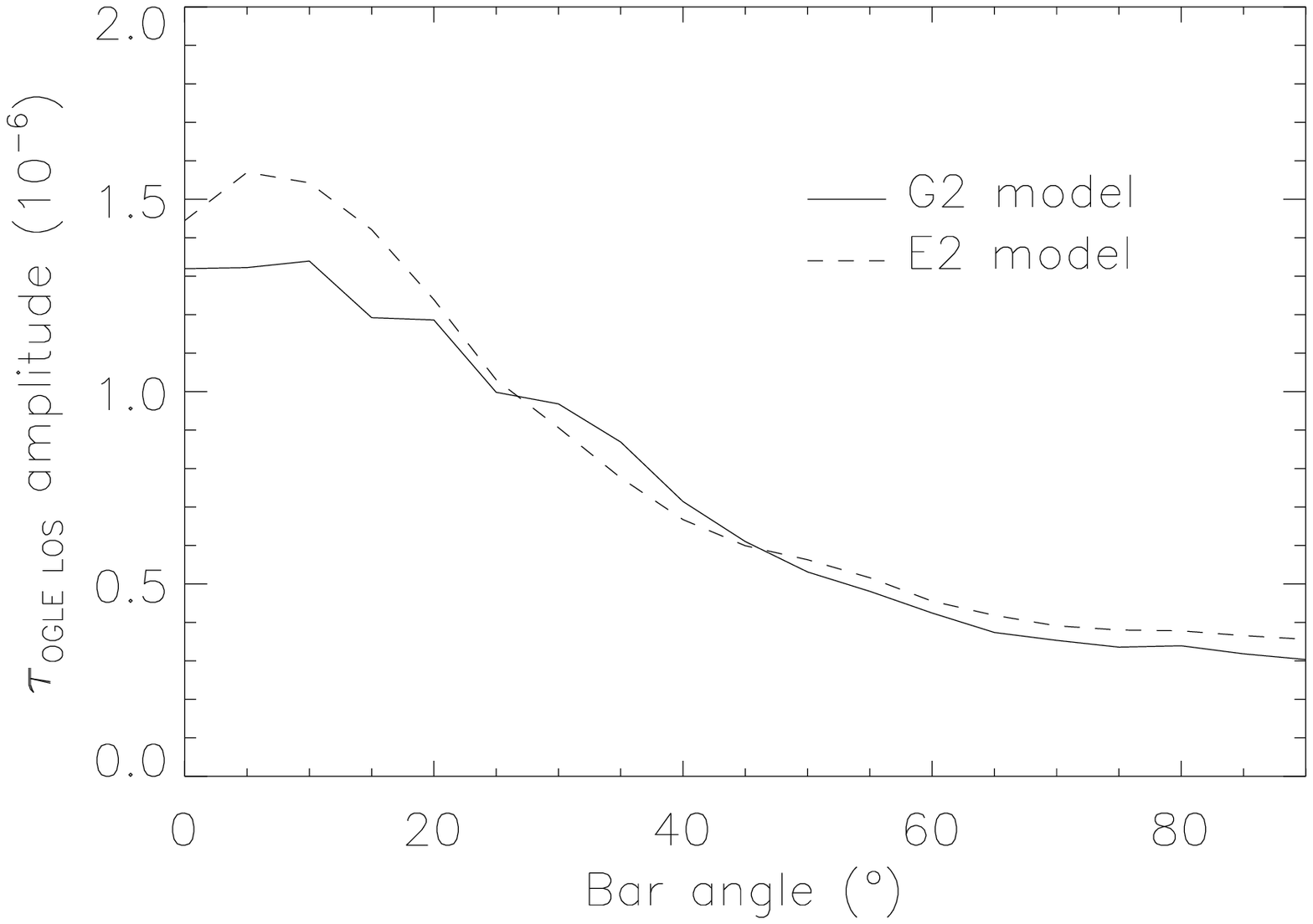}
\includegraphics[width = 7.5cm]{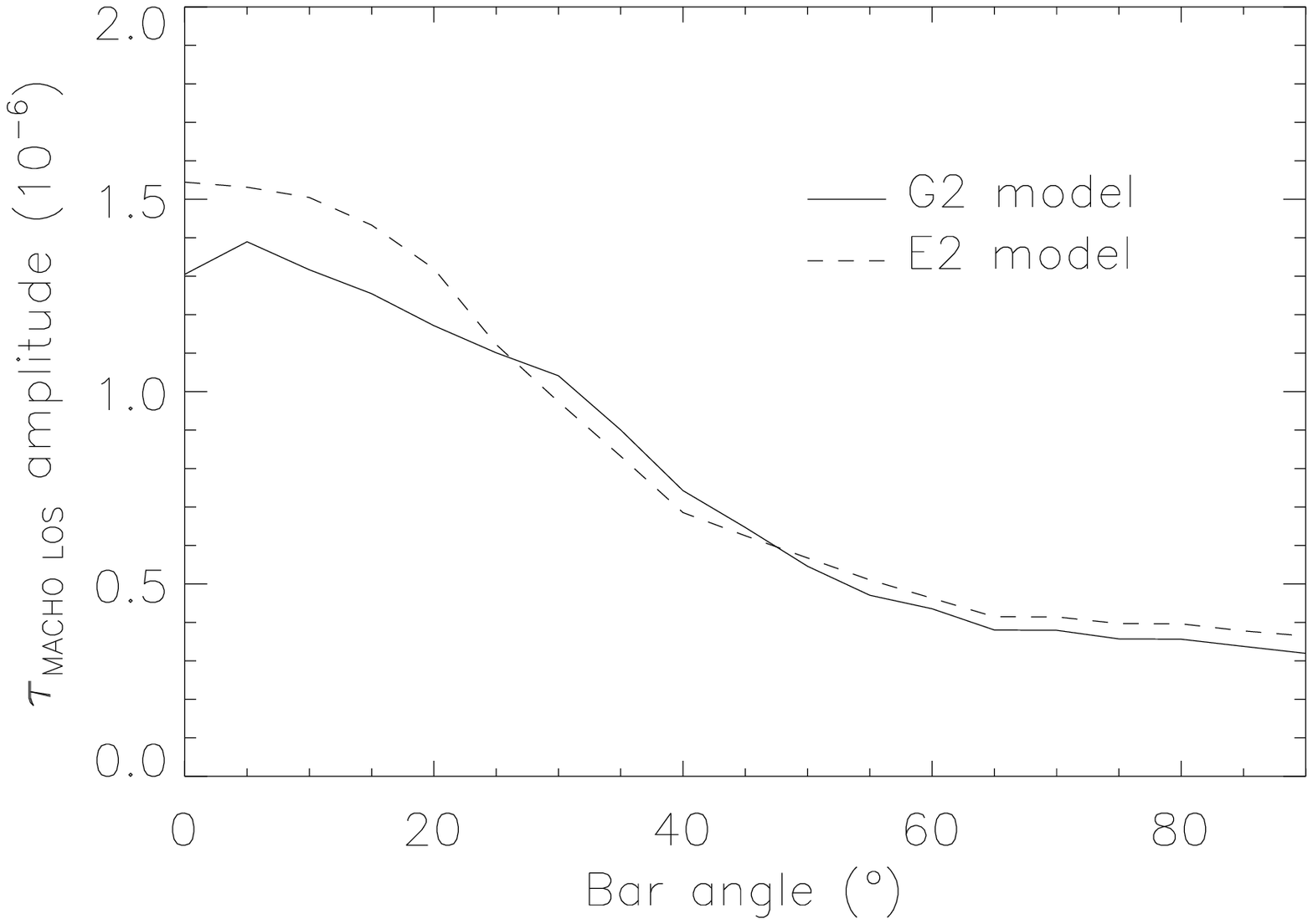}
\includegraphics[width = 7.5cm]{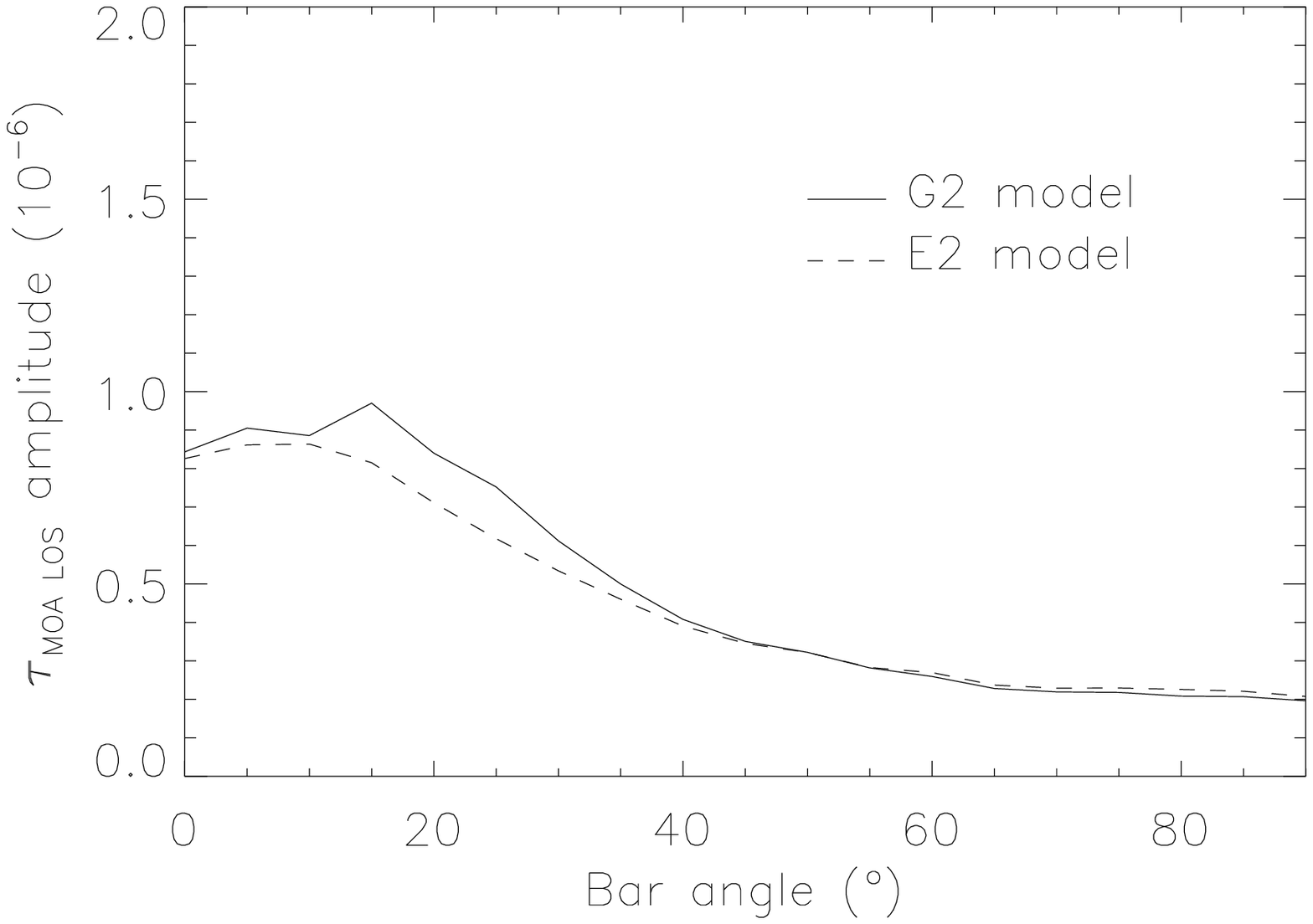}
\caption{(Top, middle, bottom) panel: oscillation amplitude of expected ($\tauoglelos$, 
  $\taumacholos$, $\taumoalos$) as a function of $\thetabar$, for the G2 and E2 models.}
\label{fig:magtau_range}
\end{figure}

\begin{figure}
\centering
\includegraphics[width = 7.5cm]{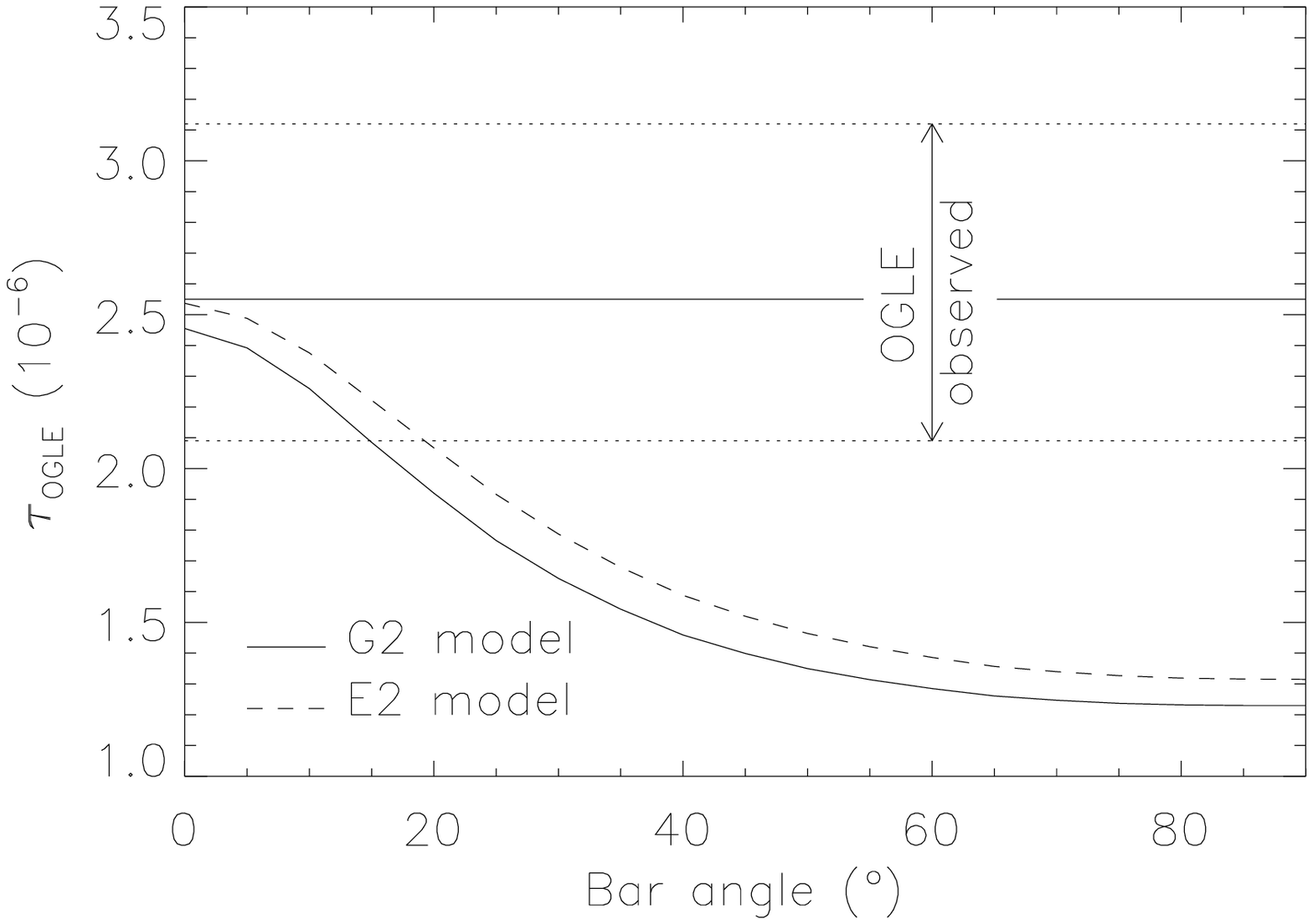}
\includegraphics[width = 7.5cm]{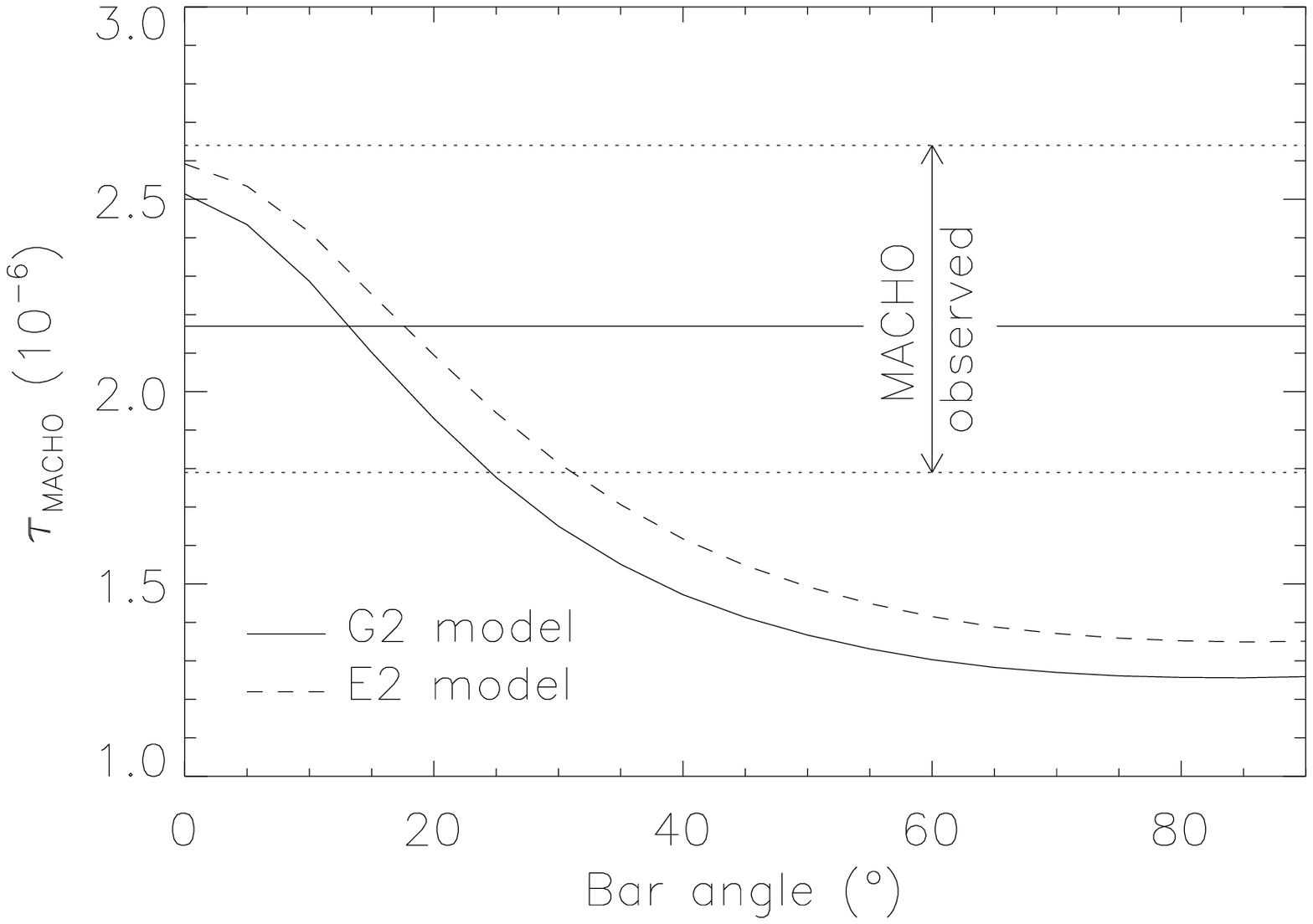}
\includegraphics[width = 7.5cm]{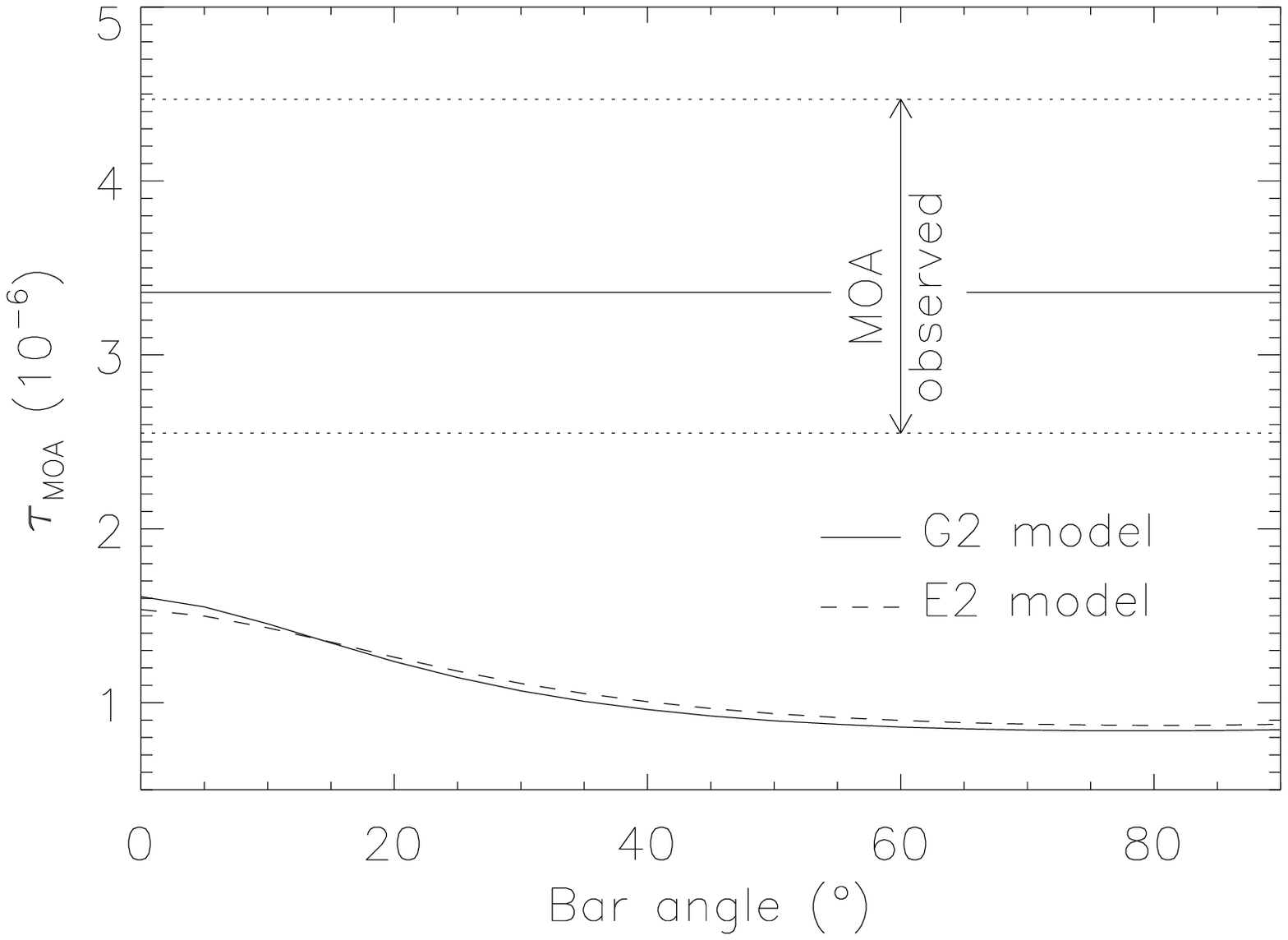}
\caption{(Top, middle, bottom) panel: expected ($\tauogle$, $\taumacho$, $\taumoa$) as 
  a function of $\thetabar$, for the G2 and E2 models. The horizontal lines show the 
  value measured by (OGLE, MACHO, MOA) with its $1 \sigma$ uncertainties.}
\label{fig:angtau}
\end{figure}

If the given bar angles of the G2 and E2 models are now varied, model-independent 
trends with $\thetabar$ may be revealed. Fig. \ref{fig:magtau_range} shows that the 
expected amplitudes of $\tauoglelos$, $\taumacholos$ and $\taumoalos$ all decrease 
with increasing bar angle. This provides a potential constraint on $\thetabar$, 
should the expected $\tau$ amplitude be observed and its magnitude accurately measured. 
Another constraint is shown by Fig. \ref{fig:angtau}, where the expected optical depths 
$\tauogle$, $\taumacho$ and $\taumoa$ display a similar dependance on $\thetabar$. The 
corresponding observed values are overplotted, with their $1 \sigma$ uncertainties, and 
from the intersections with the predicted OGLE and MACHO curves, $1 \sigma$ upper 
limits on $\thetabar$ are obtained. (There is no intersection between the predicted and 
observed $\taumoa$). These limits are given in Table \ref{tab:thetabar_limits}. Note 
that they exclude the large bar angle of the E2 model ($\thetabar = 41.3^\circ$), as 
well as those from GLIMPSE ($\thetabar = (44 \pm 10)^\circ$) and EROS 
($\thetabar = (49 \pm 8)^\circ$).

\subsection{Comparison with EROS data}
\label{sec:results_eros}

The EROS-2 survey (\citealt{Ham06}) has found the largest sample of clump--giant 
events so far, 120, compared with 32 and 62 for the latest OGLE and MACHO surveys, 
respectively. This sample may be sufficient to enable a useful comparison of the 
predicted optical depth trends with observational data. Since the EROS-2 survey (like 
all microlensing surveys) was conducted across many fields rather than for a specific 
LOS, any observed oscillation in $\tau$ similar to the prediction would be somewhat 
smoothed out. However, this effect should not be strong, as the predicted trend is 
similar for different lines of sight towards the bulge (as shown in Fig. 
\ref{fig:magtau}), and in any case such effects are accounted for as described below.

\begin{figure}
\centering
\includegraphics[width = 7.5cm]{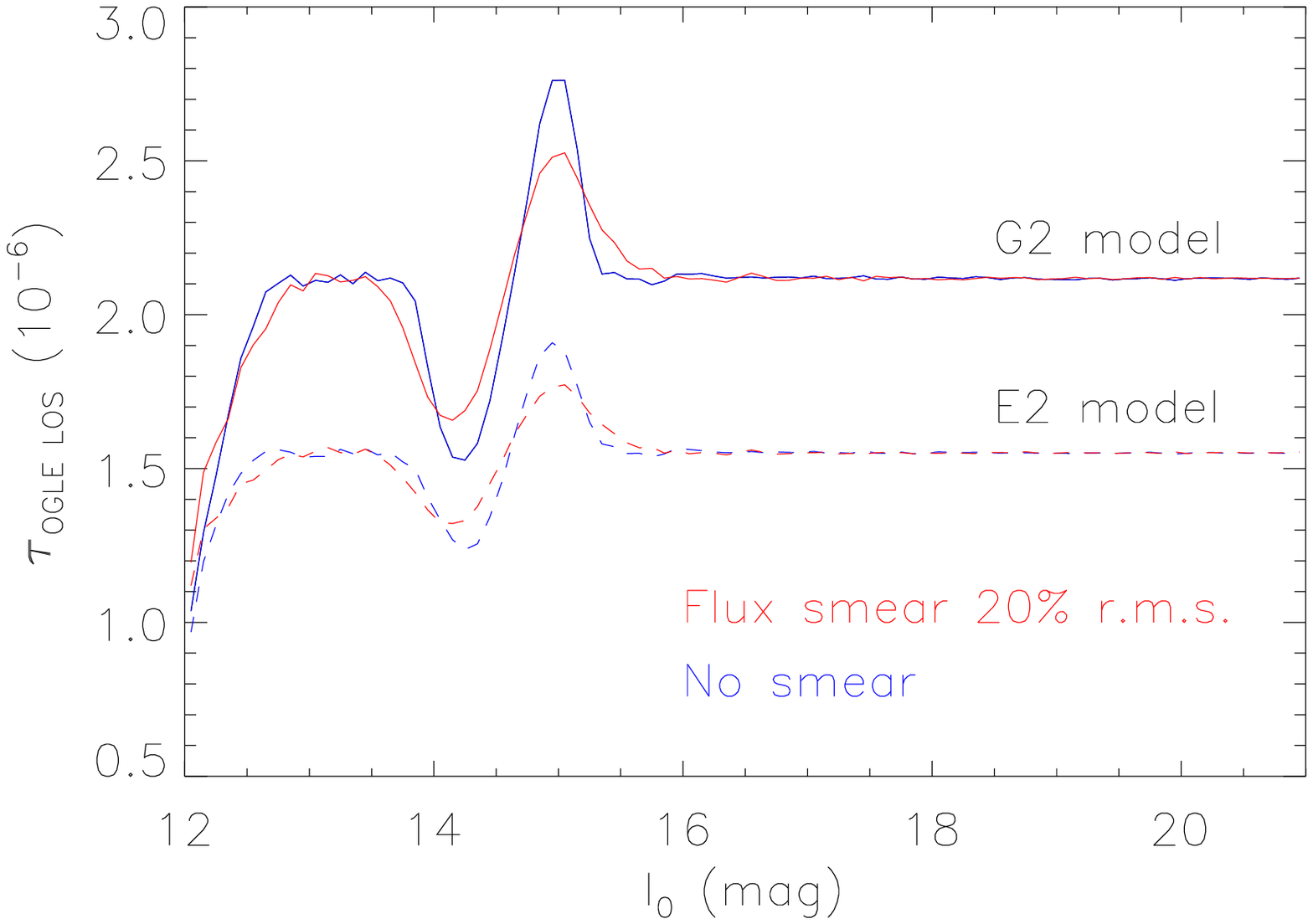}
\includegraphics[width = 7.5cm]{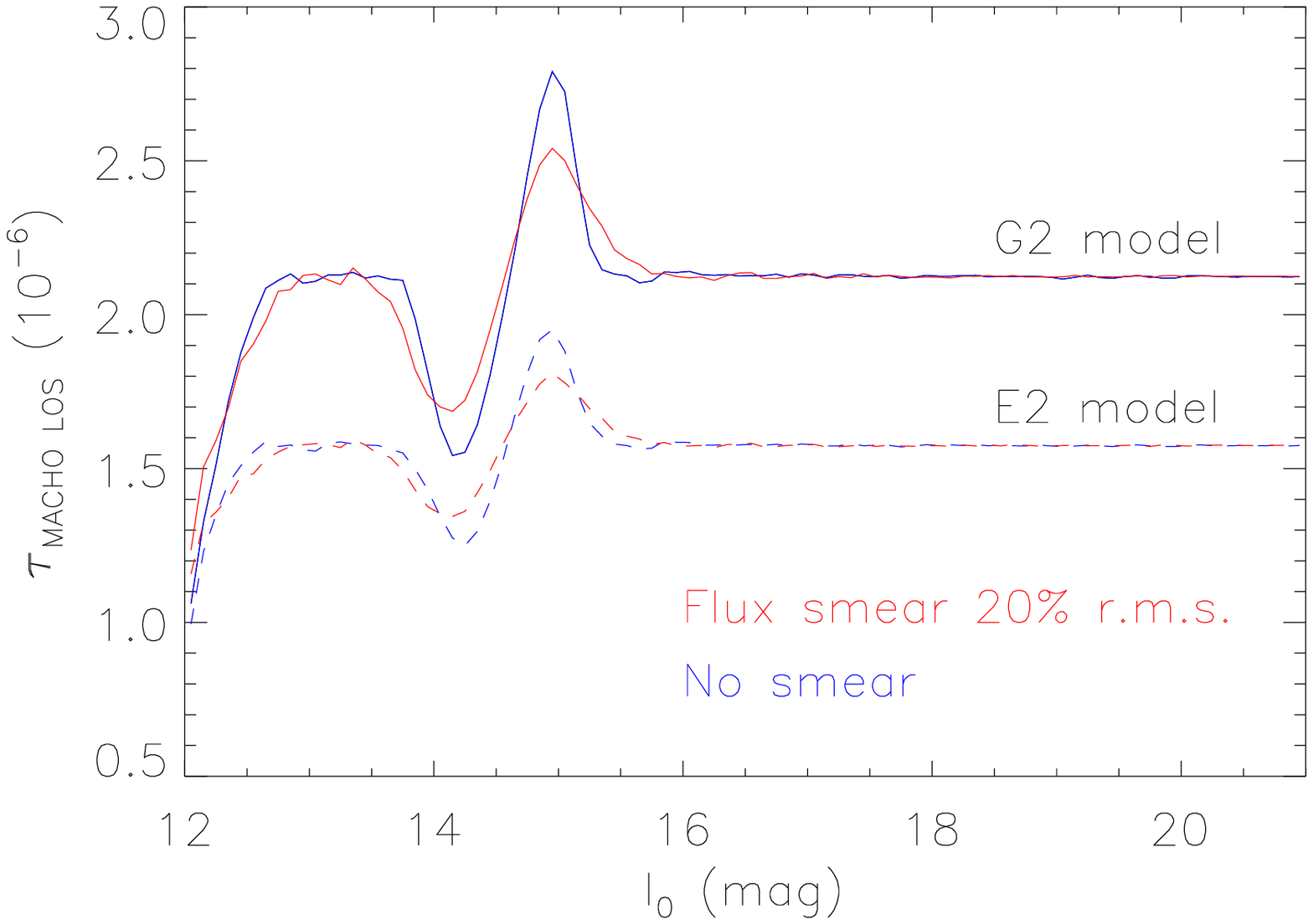}
\includegraphics[width = 7.5cm]{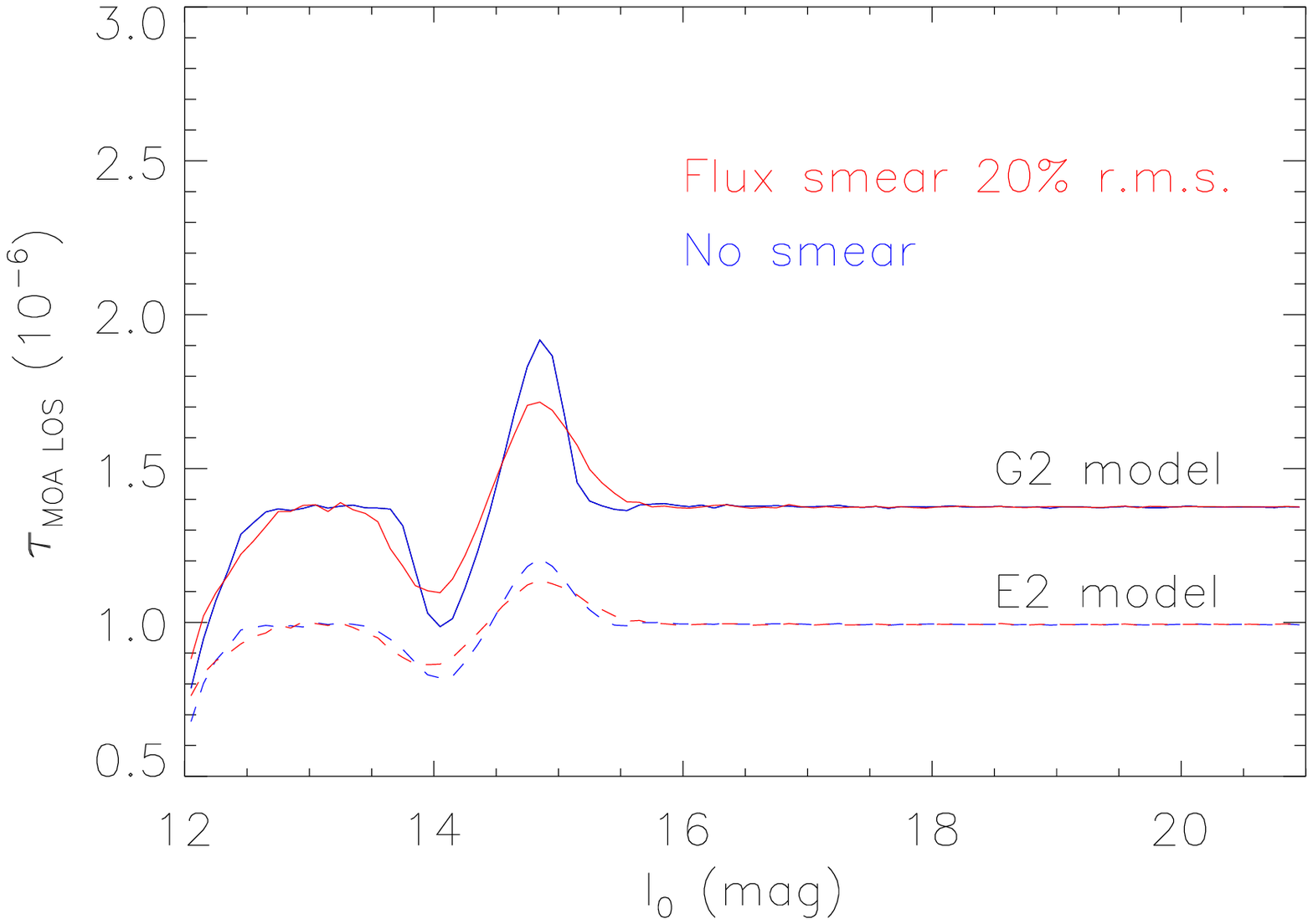}
\caption{(Top, middle, bottom) panel: expected ($\tauoglelos$, $\taumacholos$, 
  $\taumoalos$) as a function of source apparent magnitude, with and without stellar 
  flux smearing of 20 per cent rms, as indicated. The curves with no smearing are the 
  same as in Fig. \ref{fig:magtau}.}
\label{fig:magtau_smear}
\end{figure}

To make the comparison, unpublished EROS-2 data have been supplied by J. Rich et al. 
\citetext{private communication}. They find, from studies with artificial stars, that 
clump giant fluxes are smeared by $\sim$20 per cent rms, which does not affect their 
optical depth calculations averaged over the whole clump, but will of course reduce 
the $\tau$ amplitude by smoothing out the predicted oscillation. This flux smearing 
effect is added to all of the model stars, using a Gaussian with $\sigma$ = 0.2. It 
is found that although the $\tau$ amplitude is indeed reduced, the oscillation is 
still clear, as shown in Fig. \ref{fig:magtau_smear}.

EROS stars were observed in two non-standard bands, $\Reros$ and $B_{\rm EROS}$, 
where $\Reros = \Iogle$. Hence only the $\Reros$ data are considered, to avoid any 
complications arising from using different magnitude scales. \citet{Ham06} divided 
each of their 66 bulge fields into 32 subfields. For each subfield, they modelled 
the stellar density in colour--magnitude space as a power-law plus Gaussian, in order 
to find the magnitude, $\Rclump$, and colour of the clump centre. Hence smoothing 
effects (as mentioned above) and extinction can now both be easily accounted for here, 
by taking the clump centres as reference points: instead of simply finding $\tau$ as a 
function of $\Reros$, it is found as a function of magnitude \emph{relative to the 
clump centres}, $\Reros - \Rclump$. This method `lines up' all the clump centres in 
each subfield. For comparison, the predicted $\tau$ is similarly found as a function 
of $I_0 - \Iclump$. For the (OGLE, MACHO, MOA) LOS, the Gaussian component shown in 
Fig. \ref{fig:freq_sources} peaks at $\Iclump$ = (14.75, 14.65, 14.55) $\pm\ 0.05$ (for 
bin widths of 0.1 mag). For the E2 model, the magnitudes are slightly brighter: 
$\Iclump$ = (14.65, 14.55, 14.45) $\pm\ 0.05$.

$\taueros$ is now found as a function of source magnitude using equation (13) from 
\citet{Ham06}:
\begin{eqnarray}
\tau & = & {\pi \over 2 u_0 (\rm max)} {\sum_{i = 1}^{N_{\rm ev}} t_{{\rm E}, i} / \epsilon (t_{{\rm E}, i}) \over \sum_{j = 1}^{N_*} T_j},
\end{eqnarray}
where each event $i$ has a time-scale $t_{{\rm E}, i}$ and detection efficiency 
$\epsilon (t_{{\rm E}, i})$, each monitored star $j$ is observed for a time $T_j$, the 
total numbers of events and stars are $N_{\rm ev}$ and $N_*$, respectively, and the 
maximum impact parameter $u_0 (\rm max)$ = 0.75. This calculation is implemented with a 
separate summation for each magnitude bin (and taking full account of the different 
detection efficiencies for each EROS-2 field). The uncertainty is also determined 
following \citet{Ham06}, who added in quadrature a 5 per cent systematic part -- due to 
blending effects -- and a larger statistical part, estimated according to \citet{HG95}. 

\begin{figure}
\centering
\includegraphics[width = 7.5cm]{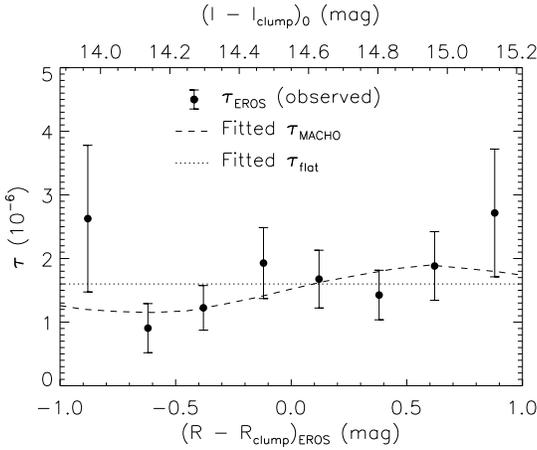}
\caption{Observed $\taueros$, and fitted model $\taumacholos$, as functions of 
  magnitude relative to the clump centre, $\Reros - \Rclump$ and $I_0 - \Iclump$, 
  respectively (see text). The latter magnitude scale is stretched relative to the 
  former by the fitting factor. Also shown is the fitted $\tauflat$.}
\label{fig:magtau_eros}
\end{figure}

Fig. \ref{fig:magtau_eros} shows $\taueros$ as a function of $\Reros - \Rclump$. The 
observed trend is now compared with the flux-smeared predicted trends, using a $\chi^2$ 
test, to see if the former is better fitted by an oscillating or constant optical depth. 
Since it is fitting an oscillating \emph{form} that is of interest, the width of the 
oscillation and the absolute values of the $\tau$ amplitude and $\tauflat$ are allowed 
to vary as free parameters in the fit: the model trend may be stretched along the 
magnitude axis, and both shifted and stretched along the $\tau$ axis -- though not 
stretched in the former case by more than an arbitrary limit of 50 per cent, since Fig. 
\ref{fig:magtau} shows that the width of the oscillation does not vary much with 
direction. Finally, to allow for other slight changes in the shape of the oscillation 
with direction and bulge model, all six of the predicted trends (three lines of sight, 
G2 and E2 models) are tested. These $\chi^2$ values are compared with that for a 
freely-fitted constant optical depth. The results are shown in Table 
\ref{tab:chi_sq_rel}. Also indicated are the chance probabilities $p$ that $\chi^2$ 
would be greater than or equal to the given values.

An oscillating $\tau$ appears to provide a better fit to the data than a constant 
optical depth. There is a mostly negligible change in $\chi^2$ with direction, and a 
small but insignificant preference for the E2 model. Fig. \ref{fig:magtau_eros} shows 
the best-fit oscillation, with the MACHO E2 trend: $\tauflat$ has been shifted by 
$-0.05 \times 10^{-6}$, and the curve has been stretched by factors of 1.50 and 1.60 
along the magnitude and $\tau$ axes, respectively. (Note that the factor of 1.50 is at 
the (arbitrary) 50 per cent limit. The fits improve with further magnitude stretching, 
the best possible fit being for a factor of 2.50 (MACHO G2 model), with 
$\chi^2 = 1.44$, but this is well beyond the limit and is ignored). However, the 
significance of the $\chi^2$ preference for an oscillating $\tau$, rather than a 
constant value, is low. A reasonable magnitude binning gives only eight data points. 
Whereas the oscillating $\tau$ fit has three free parameters and five degrees of 
freedom, the constant $\tau$ fit has of course just one free parameter, and seven 
degrees of freedom. The constant $\tau$ fit is not significantly discrepant to the data.

It is possible that the EROS-2 event detection efficiency may be a function of 
magnitude. However, it would not be a strong dependance, and any variation would be 
smooth. It could not therefore hide any real oscillation of $\tau$, or generate a false 
one \citetext{J. Rich, private communication}. It appears that the available data are 
still insufficient to accurately determine the dependance of the optical depth on 
source apparent magnitude.

\begin{table}
\centering
\begin{tabular}{lcc}\hline

                  & G2           & E2                \\ \hline
  OGLE            & 2.07         & 1.84              \\
  MACHO           & 1.91         & 1.83              \\
  MOA             & 1.99         & 1.93              \\
                  &              &                   \\
  Average         & 1.99         & 1.87              \\
                  & ($p > 0.75$) & ($p > 0.75$)      \\
                  &              &                   \\
  Constant $\tau$ & \multicolumn{2}{c}{4.24}         \\
                  & \multicolumn{2}{c}{($p > 0.75$)} \\
                  &              &                   \\
  $\Delta \chi^2$ & 2.25         & 2.37              \\ \hline

\end{tabular}
\caption{$\chi^2$ values from fitting the observed $\taueros$ as a function of 
  $\Reros - R_{\rm clump}$ with the predicted (oscillating) trends, for different lines 
  of sight and bulge models as indicated, and with a constant optical depth. Also 
  indicated are the chance probabilities $p$ that $\chi^2$ would be greater than or 
  equal to the given values. An oscillating $\tau$ provides a better fit, but at a low 
  significance (see text).} 
\label{tab:chi_sq_rel}
\end{table}

\begin{figure}
\centering
\includegraphics[width = 7.5cm]{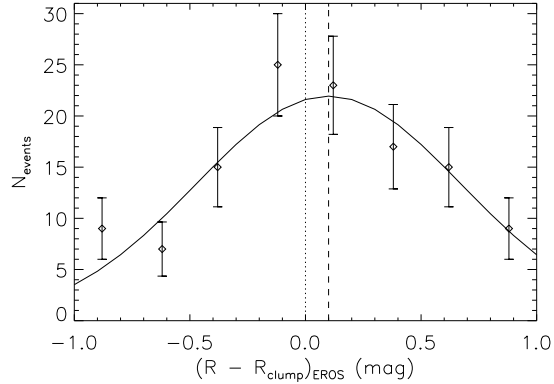}
\caption{EROS event counts as a function of magnitude relative to the clump centre, 
  $\Reros - \Rclump$. The error bars represent Poisson uncertainties. The distribution 
  is skewed towards the faint side of the magnitude axis; the trend is fitted with a 
  Gaussian in order to test the significance of the skew (see text). The vertical 
  dotted line indicates $\Reros - \Rclump = 0$, and the vertical dashed line indicates 
  the centre of the Gaussian.}
\label{fig:eros_skew}
\end{figure}

However, there is another, simpler way to look for signs of the predicted $\tau$ 
oscillation in the survey data. If $\tau$ is indeed higher on the faint side of the 
clump centres, then more of the observed events should also be on the faint side: a 
plot of event counts, as a function of $\Reros - \Rclump$, should be noticeably skewed 
towards the faint side. Fig. \ref{fig:eros_skew} shows that this is so. The ratio of 
events with $\Reros - \Rclump$ fainter than zero to those brighter than zero (hereafter 
the \emph{skewness ratio}) is 1.14. The significance of this skew is tested as follows. 
The plot of event counts is fitted with a Gaussian, as shown. This is of course an 
imperfect fit, but gives an approximate measure of the dispersion. Then 120 points (for 
the 120 EROS events) are randomly distributed on the magnitude axis, according to a 
Gaussian with the fitted dispersion, and the skewness ratio is calculated. For one 
million such Monte Carlo cases, only 21 per cent have a skewness ratio greater than the 
observed value. Thus the observed skew is suggestive, but not highly significant.

\section{Summary and conclusions}
\label{sec:conclusions}

It does not appear that the discrepancy in optical depth measurements between the RCG 
and all-star analyses can be explained by a dependence of the lensing surveys on their 
flux limits. The model reproduces the OGLE and MACHO values based on RCGs, but 
underpredicts MOA's all-star value by $\sim$$2.4 \sigma$. Another potential explanation 
for the discrepancy is blending. \citet{Sum06} found $\sim$38 per cent of OGLE-II 
events with apparent RCG sources were really due to faint stars blended with a bright 
companion. However, they also showed that blending has little effect on estimates of 
$\tau$ due to partial cancellation of its different effects, a point also made by 
\citet{Pop05} and \citet{Ham06}. \citet{Sum06} also state that the DIA method is less 
sensitive to the systematics of blending in crowded fields. Though it is of course 
possible that MOA's value may yet be lowered with more data, it is supported by MACHO's 
earlier DIA value.

$\tau$ is expected to be generally constant as a function of source apparent magnitude 
for $I_0 \gtrsim 13.0$, except in the range $13.5 \lesssim I_0 \lesssim 15.5$, where 
many RCGs are detected. These stars dominate the source counts at such magnitudes, and 
show a strong correlation between distance and apparent magnitude, causing a 
significant oscillation in $\tau$. The amplitude of this oscillation is found to 
decrease with increasing bar angle, providing a potential constraint on $\thetabar$. A 
further constraint comes from a similar dependance of $\tau$ with $\thetabar$: 
combining the predicted trends with the measured values provides $1 \sigma$ upper 
limits, which exclude the large bar angles reported by GLIMPSE and EROS.

From the EROS-2 survey, $\taueros$ has been found as a function of source apparent 
magnitude. The predicted oscillation is not only consistent with the observed trend, 
but provides a better $\chi^2$ fit than a constant optical depth, though the 
significance of this preference is low due to insufficient data. However, a further 
sign comes from EROS event counts, which show a clear skew towards fainter magnitudes. 
With ongoing surveys detecting increasing numbers of RCG events (and $\sim$$500\ 
{\rm yr}^{-1}$ of all kinds), it should soon be possible to make a more useful and 
definite comparison.

\section*{Acknowledgments}

I am grateful to Jim Rich of the EROS collaboration for providing their data, and for 
his helpful comments. I also thank Szymon Koz{\l}owski, Shude Mao and {\L}ukasz 
Wyrzykowski for useful discussions, and the anonymous referee for their comments. I 
acknowledge support from a PPARC studentship, and travel support from the European 
Community's Sixth Framework Marie Curie Research Training Network Programme, Contract 
No. MRTN-CT-2004-505183 `ANGLES'.

\newpage

\bibliographystyle{mn2e}

\begin{thebibliography}{}

\bibitem[\protect\citeauthoryear{Afonso et al.}{2003}]{Afo03} 
         Afonso C. et al., 2003, \aap, 404, 145
\bibitem[\protect\citeauthoryear{Alcock et al.}{2000}]{Alc00} 
         Alcock C. et al., 2000, \apj, 541, 734
\bibitem[\protect\citeauthoryear{Benjamin et al.}{2005}]{Ben05} 
         Benjamin R.A. et al., 2005, \apj, 630, L149
\bibitem[\protect\citeauthoryear{Dwek et al.}{1995}]{Dwe95} 
         Dwek E. et al., 1995, \apj, 445, 716
\bibitem[\protect\citeauthoryear{Gerhard}{2002}]{Ger02} 
         Gerhard O., 2002, in Da Costa G.S., Sadler E.M., eds., ASP Conf. Ser. Vol. 
         273, The Dynamics, Structure \& History of Galaxies: A Workshop in Honour of 
         Professor Ken Freeman, Astron. Soc. Pac., San Francisco, p. 73
\bibitem[\protect\citeauthoryear{Hamadache et al.}{2006}]{Ham06} 
         Hamadache C. et al., 2006, \aap, 454, 185
\bibitem[\protect\citeauthoryear{Han \& Gould}{1995}]{HG95} 
         Han C., Gould A., 1995, \apj, 449, 521
\bibitem[\protect\citeauthoryear{Han \& Gould}{2003}]{HG03} 
         Han C., Gould A., 2003, \apj, 592, 172
\bibitem[\protect\citeauthoryear{Popowski et al.}{2005}]{Pop05} 
         Popowski P. et al., 2005, \apj, 631, 879
\bibitem[\protect\citeauthoryear{Stanek}{1995}]{Sta95} 
         Stanek K.Z., 1995, \apj, 441, L29
\bibitem[\protect\citeauthoryear{Stanek et al.}{1994}]{Sta94} 
         Stanek K.Z., Mateo M., Udalski A., Szyma\'{n}ski M., Kalu\.{z}ny J., 
         Kubiak M., 1994, \apj, 429, L73
\bibitem[\protect\citeauthoryear{Sumi}{2004}]{Sum04} 
         Sumi T., 2004, \mnras, 349, 193
\bibitem[\protect\citeauthoryear{Sumi et al.}{2003}]{Sum03} 
         Sumi T. et al., 2003, \apj, 591, 204
\bibitem[\protect\citeauthoryear{Sumi et al.}{2006}]{Sum06} 
         Sumi T. et al., 2006, \apj, 636, 240
\bibitem[\protect\citeauthoryear{Thomas et al.}{2005}]{Tho05} 
         Thomas C.L. et al., 2005, \apj, 631, 906
\bibitem[\protect\citeauthoryear{Udalski et al.}{2002}]{Uda02} 
         Udalski A. et al., 2002, Acta Astron., 52, 217
\bibitem[\protect\citeauthoryear{Wood \& Mao}{2005}]{WM05} 
         Wood A., Mao S., 2005, \mnras, 362, 945
\bibitem[\protect\citeauthoryear{Wo\'{z}niak et al.}{2001}]{Woz01} 
         Wo\'{z}niak P.R., Udalski A., Szyma\'{n}ski M., Kubiak M., Pietrzy\'{n}ski G., 
         Soszy\'{n}ski I., \.{Z}ebru\'{n} K., 2001, 
         Acta Astron., 51, 175
\bibitem[\protect\citeauthoryear{Zheng et al.}{2001}]{Zhe01} 
         Zheng Z., Flynn C., Gould A., Bahcall J.N., Salim S., 2001, \apj, 555, 393

\end{thebibliography}

\end{document}